\documentclass[useAMS,usenatbib]{mn2e}
\usepackage{soul}
\usepackage{amssymb}
\usepackage{amsmath}
\usepackage{float}
\usepackage{times}
\usepackage[usenames,dvipsnames]{color}
\usepackage{graphicx,epsfig,graphics}
\usepackage{multirow}

\newcommand{\san}[1]{\textcolor{black}{ #1}}
\title[Radio Interferometric Calibration Using The SAGE Algorithm]{Radio Interferometric Calibration Using The SAGE Algorithm}

\author[Kazemi et al. ]{ S. Kazemi$^{1}$\thanks{E-mail:
kazemi@astro.rug.nl}, S. Yatawatta$^{1,2}$, S. Zaroubi$^{1,3}$, A.G. de Bruyn$^{1,2}$,  L.V.E. Koopmans$^{1}$, 
\newauthor J. Noordam$^{2}$ \\ $^{1}$Kapteyn Astronomical Institute, University
of Groningen, P.O. Box 800, 9700 AV Groningen, the Netherlands\\
$^{2}$ASTRON, Postbus 2, 7990 AA Dwingeloo, \\The Netherlands 
$^{3}$Physics Department, Technion, Haifa 32000, Israel}

\begin{document}

% \date{Accepted 20XX Month XX. Received 20XX Month XX; in original form
% 20XX Month XX}

\pagerange{\pageref{firstpage}--\pageref{lastpage}} \pubyear{2007}

\maketitle 

\label{firstpage}

\begin{abstract}
The aim of the new generation of radio synthesis arrays such as LOFAR and SKA is to achieve much higher sensitivity, resolution and frequency coverage than what is available now, \san{especially at low frequencies}. To accomplish this goal, the accuracy of the calibration techniques used is of considerable importance. Moreover, since these telescopes produce huge amounts of data, speed of convergence of calibration is a major bottleneck. The errors in calibration are due to system noise (sky and instrumental) as well as the estimation errors introduced by the calibration technique itself, which we call ``solver noise''. We define solver noise as the ``distance'' between the optimal solution \san{(the true value of the unknowns, uncorrupted by the system noise)} and the solution obtained by calibration. We present the Space Alternating Generalized Expectation Maximization (SAGE) calibration technique, which is a modification of the Expectation Maximization algorithm, and compare its performance with the traditional Least Squares calibration based on the level of solver noise introduced by each technique. For this purpose, we develop  statistical methods that use   the calibrated solutions to estimate the level of solver noise. The SAGE calibration algorithm yields very promising results both in terms of accuracy and speed of convergence. The comparison approaches that we adopt introduce a new framework for assessing the performance of different calibration schemes.
\end{abstract}


\section{Introduction}
\label{sec:introduction}

Early radio-astronomy predominantly used single-dishes for observations. With the resolution requirements increasing, the single dish approach became impractical. This paved the path for using radio-interferometric techniques with multiple  antennas linked together as an array  that operates as a large effective single-dish \citep{A.R.1}.

The sensitivity of an interferometer is greatly increased, compared to a single-dish telescope, due to the larger combined collecting area. The currently planned or built  radio interferometers, such as the \san{Square Kilometer Array (SKA)\footnote{http://www.skatelescope.org}, the Murchison Widefield Array (MWA)\footnote{http://www.mwatelescope.org}, the Precision Array to Probe Epoch of Reionization (PAPER)\footnote{http://astro.berkeley.edu/\~{}dbacker/eor}, the 21-cm Array (21CMA)\footnote{http://21cma.bao.ac.cn}, the Hydrogen Epoch of Reionization Array (HERA)\footnote{http://www.reionization.org}, the Long Wavelength Array (LWA)\footnote{http://lwa.unm.edu}  and the LOw Frequency ARray (LOFAR)\footnote{http://www.lofar.org}, consist of a large number of elements and include short, intermediate and many of them longer antenna spacings}. For an introduction to radio interferometry we refer the reader to \cite{A.R.1}.

In the interferometric visibilities there always exist errors introduced by the sky, the atmosphere (e.g. troposphere and ionosphere), the instrument (e.g. beam-shape, frequency response, receiver gains etc.) and by Radio Frequency Interference (RFI). The process of estimating and reducing the errors in these measurements is called ``calibration'' and is an essential step before imaging the visibilities.

The classical calibration method, named external (or primary) calibration, is based on observing a celestial radio source with known properties. This approach is strongly dependent on the accuracy with which the source properties are known. The external calibration is improved by using self-calibration \citep{selfcal} which utilizes  the observed data for estimating both the unknown instrumental and the sky parameters. The quality of calibration and the imaging is significantly increased by iterating between the sky and the instrument model. \san{The redundant calibration is also independent of the sky model. It calibrates for both the sky and the instrument, using redundant information in the measured data. However, its performance is limited to arrays with a regular arrangement in their antennas layout.}

Calibration is an optimization process that is non-linear by nature. It is in essence a Maximum Likelihood (ML) estimation of the unknown parameters by applying non-linear optimization techniques. Traditional calibration is estimating the ML solution by the non-linear Least Squares (LS) method via various gradient-based techniques such as the Levenberg-Marquardt (LM) algorithm \citep{K.L1, A.L.1}. This approach was improved by the Expectation Maximization (EM) algorithm \citep{M.2} and later on by the Space Alternating Generalized Expectation Maximization (SAGE) technique which was introduced by \citet{J.A.1} and was applied to interferometer calibration by \citet{S.2}. The analysis and application of the aforementioned schemes for the  calibration of radio interferometers can be found in \citet{S.2}.  

To reach the scientific goals of the new generation of radio arrays, calibration algorithms must have the highest accuracy possible. Furthermore, the number of measured visibilities that has to be calibrated  is unprecedented. The speed of convergence of the calibration processes must therefore be the fastest with the minimum possible computational cost. Based on these facts, the best calibration method is referred to as the one which minimizes the ``distance'' between the true values of unknown parameters and the values obtained by calibration and minimizes computational time.

We should take into account that the measured data of an interferometer is always corrupted by different sources of noise such as the thermal noise, which is an additive Gaussian random process, and confusion noise \citep{condon}, which affects the coherency matrix (see next section and e.g. \citet{bornwolf}). For a detailed discussion on the sources of noise the reader is referred to the Chapter 6 of \citet{stef}. When the calibration process of the measured data is done, the ``distance'' between the true value of the unknown parameters and their calibrated solutions depends on the initial noise and the errors originating from the calibration process itself (e.g. converging to a local minimum), which is called ``solver noise''. In other words, because the calibrated solutions are not optimal, there always exists some solver noise between these solutions and the true values of the unknown parameters affected by the initial noise. The lower the solver noise, the higher the accuracy of the calibrated results. Thus, in order to increase the calibration efficiency, we need to choose the calibration scheme which has the minimum solver noise as well as the lowest computational cost. To achieve this, we should be able to compare these two factors between various calibration techniques. We introduce a general framework for detecting  the level of solver noise in calibration algorithms based only on their solutions.

In this paper, we present the SAGE calibration method and emphasize its superiority, compared to the traditional LS calibration, in terms of accuracy and speed of convergence. Mathematical derivations of the algorithms are presented in the appendices. We also investigate the applicability of two well-known measures, the Kullback-Leibler Divergence (KLD) \citep{K.L.D} and the Likelihood Ratio Test (LRT) \citep{L.R.T}, in  revealing  the level of solver noise in calibrated solutions. Illustrative examples of both real and simulated observations, show the superior performance of the SAGE calibration compared to the LS one. \san{They also indicate that the LRT approach is very promising at detecting the level of solver noise in the obtained calibrated solutions, while the KLD approach is not always conclusive.}

The following notations are used in this paper: Bold, lowercase letters refer to column vectors, e.g., {\bf y}. Upper case bold letters refer to matrices, e.g., {\bf C}.  All parameters are complex numbers, unless stated otherwise. The inverse, transpose, Hermitian transpose, and conjugation of a matrix are presented by $(.)^{-1}$, $(.)^T$, $(.)^H$, and $(.)^*$, respectively. The statistical expectation operator is referred to as $E\{.\}$. The matrix Kronecker product and the proper (strict) subset are denoted by $\otimes$ and $\subsetneq$, respectively. The diagonal matrix consisting of only the diagonal entries of a square matrix is given by diag$(.)$. $\bf{I}$ is the identity matrix and $\varnothing$ is the empty set. The Kronecker delta function is presented by $\delta_{ij}$. $\mathbb{R}$ and $\mathbb{C}$ are the sets of Real and Complex numbers, respectively. The Frobenius norm is shown by $||.||$. Estimated parameters are denoted by a hat, $\widehat{(.)}$. All logarithmic calculations are to the base $e$.

\section{The measurement equation}\label{modeling}
  
The first stage in the (self-)calibration process is to provide an efficient measurement equation which relates the visibilities with the unknown sky and the instrument parameters. In this section, we use the measurement equation presented by \citet{J.P.1}. For assessing the equation from the array signal processing point of view, the reader is referred to \citep{A.J.1, 2003ITSP...51...25B, S.1}.

We assume that we have a radio interferometer consisting of $N$ receiver antennas. Each antenna consists of two orthogonal dual-polarization feeds, which receive the incident polarized waves from astrophysical sources in the sky. We also assume that the radio frequency sky consists of $K$ discrete, \san{uncorrelated} sources. The sources are far enough from the array that their radiations can be assumed to be plane waves.
  
Let us consider ${\bf{e}}_i=[e_{Xi}\ e_{Yi}]^T$ represents the electric field vector of the $i$-th source. This field causes an induced voltage $\tilde{{{\bf{v}}}}_{pi}=[v_{Xpi}\ v_{Ypi}]^T$ at \san{antenna} $p$ for every $p\in\{1,2,\ldots,N\}$ due to:
\begin{equation}
\tilde{{{\bf{v}}}}_{pi}={{\bf{J}}}_{pi}{{\bf{e}}}_{i}.\label{s1}
\end{equation}
In Eq. (\ref{s1}), the $2\times 2$ Jones matrix ${{\bf{J}}}_{pi}$ describes the complex interaction between the fields, the \san{antenna}  beam-shape and ionosphere,  as well as the remaining signal path. The total signal at the \san{antenna} $p$, ${\bf v}_p$, is a linear superposition of $K$ such signals as in (\ref{s1}). At the end, the receiver noise ${\bf {\nu}}=[\nu_X\ \nu_Y]^T$ also is added to this signal.

Before correlating the voltages of the interferometer antennas, each voltage is corrected for a geometric delay depending on the location of its receiver antenna on the earth. Thereafter, the $p$-th \san{antenna} voltage gets correlated to the other $N-1$ array \san{antenna} voltages in the array correlator. The correlated voltages, referred to as {\em {visibilities}} \citep{J.P.1} of the baseline $pq$ corresponding to the $p$-th and the $q$-th antennas, $\operatorname{E}\{{\bf v}_p\otimes{\bf v}_q^H\}$, can be given as
\begin{equation}
{\bf V}_{pq}=\sum_{i=1}^K {{\bf{J}}}_{pi}({\pmb{\theta}}){{\bf{C}}}_{i}{{\bf{J}}}^H_{qi}({\pmb{\theta}})+{\bf{N}}_{pq},\ \ p,q\in \{1,2,\ldots,N\}.\label{s2}
\end{equation}
In Eq. (\ref{s2}), the Jones matrices \citep{J.P.1}, ${{\bf{J}}}_{pi}({\pmb{\theta}})$ and ${{\bf{J}}}_{qi}({\pmb{\theta}})$ describe the electromagnetic interaction of the source $i$ at \san{antennas} $p$ and $q$, respectively \citep{bornwolf}. In particular, the instrumental properties (the beam shape, low-noise amplifier gain, system frequency response, etc.) as well as the propagation properties (tropospheric and ionospheric distortions, etc.) are represented by the Jones matrix formalism. They can be considered as the direction-dependent gains of the corresponding \san{antennas} for the $i$-th source. The unknown parameter vector ${\pmb{\theta}}\in \mathbb{C}^P$ contains the parameters of \san{both} the instrument and the sky model. ${\bf N}_{pq}$ is the $2\times 2$ noise matrix of the baseline $pq$. The {\em {coherency}} matrix \citep{bornwolf, J.P.1} is defined as
\begin{equation}
{{\bf{C}}}_{i}=\operatorname{E}\{{{\bf{e}}}_{i}\otimes{{\bf{e}}}_{i}^H\},
\end{equation}
which provides us information about the polarization state of the radiation of the $i$-th source. \san{We assume that an initial estimate of the coherency matrix  ${{\bf{C}}}_{i}$ is known based on some prior information, about the source or sky properties, obtained by previous observations.} 

Calibration is essentially a step to estimate the elements of the unknown parameter vector  ${\pmb{\theta}}$, i.e., $P$ complex values or $2P$ real values. \san{The parameter vector ${\pmb{\theta}}$ is a random variable, that varies as a function of time and frequency, by nature. We assume without loss of generality that we are calibrating on a small enough time and frequency interval during which the variation of ${\pmb{\theta}}$ is negligible. We then split the integration time to several sub-intervals and apply the calibration process to all of them separately.} 

Finally, the vectorized form of Eq. (\ref{s2}) can also be written as 
\begin{equation}
{\bf v}_{pq}\equiv\mbox{vec}({\bf V}_{pq})=\sum_{i=1}^K {\bf J}^*_{qi}(\pmb{\theta})\otimes{\bf J}_{pi}(\pmb{\theta})\mbox{vec}({\bf C}_{i})+{\bf n}_{pq},\label{s3}
\end{equation}
where ${\bf n}_{pq}=\mbox{vec}({\bf N}_{pq})$. Ignoring the auto-correlations for $p=q$, and stacking up all cross correlations as ${\bf y}=[{\bf v}^T_{12}\ {\bf v}^T_{13}\ \ldots\ {\bf v}^T_{(N-1)N}]^T,$ and all noise vectors as ${\bf n}=[{\bf n}^T_{12}\ {\bf n}^T_{13}\ \ldots\ {\bf n}^T_{(N-1)N}]^T,$ we obtain the general form of the measurement equation as
\begin{equation}
{\bf y}=\sum_{i=1}^K {\bf s}_i({\pmb{\theta}})+{\bf n}.\label{s4}
\end{equation}
In Eq. (\ref{s4}), ${\bf y},{\bf n}\in \mathbb{C}^M$,where $M$ is at most $2N(N-1)$ providing all the cross-correlations. The dimension of the parameter vector  ${\pmb{\theta}}$, $P$, is a multiple of $KN$. Thus, for a large enough $N$ and a small enough $K$, there are enough constraints for estimating  ${\pmb{\theta}}$. However, when the number of sources in the sky is uncertain, the optimal $K$ could be selected using Aikaike's Information Criterion (AIC, \citet{H.1}) as presented in \citet{S.2}. The nonlinear function ${\bf s}_i(\pmb{\theta})$, defined for $i\in\{1,2,\ldots,K\}$ as
\begin{equation*}
{\bf s}_i(\pmb{\theta})\equiv\left[\begin{array}{c}
{\bf J}^*_{2i}(\pmb{\theta})\otimes{\bf J}_{1i}(\pmb{\theta})\mbox{vec}({\bf C}_{i})\\
{\bf J}^*_{3i}(\pmb{\theta})\otimes{\bf J}_{1i}(\pmb{\theta})\mbox{vec}({\bf C}_{i})\\
\vdots\\
{\bf J}^*_{Ni}(\pmb{\theta})\otimes{\bf J}_{(N-1)i}(\pmb{\theta})\mbox{vec}({\bf C}_{i})\end{array}\right],
\end{equation*}
corresponds to the contribution of the $i$-th source in the observation. The noise ${\bf n}$ is assumed to have a multivariate Gaussian distribution with zero mean and $M\times M$ covariance matrix ${\pmb{\Pi}}$, i.e., ${\bf n}\sim \mathcal{N}(0,{\pmb{\Pi}})$ \citep{S.2}. Having the measurement equation in hand, one can apply different calibration techniques for estimating the ML of the unknowns.

\section{THE LS, EM, AND THE SAGE CALIBRATION ALGORITHMS}\label{algoritms}

In this section, we briefly discuss the Least Squares (LS, Normal) calibration via the Levenberg-Marquardt (LM) algorithm. We also discuss the new robust calibration techniques, the Expectation Maximization (EM) and in particular the Space Alternating Generalized Expectation Maximization (SAGE) calibration algorithms.

\subsection{The LS calibration via LM algorithm}\label{Normalcal}
\san{LS calibration considers the additive noise $\mathcal{N}$ to be a white Gaussian noise. Because the measurement equation shown in Eq. (\ref{s4}) has the general form of a non-linear regression model \citep{A.R.,D.G.}, the likelihood of the unknown parameter ${\pmb{\theta}}$ is maximized when the sum of squared residuals is minimized. Thus, the ML estimation of ${\pmb{\theta}}$ will be equal to the below least squared error estimation}
\begin{equation}
\begin{array}{c}
\widehat{{\pmb{\theta}}}=\mbox{arg}\ \mbox{min}\ ||{\bf y}-\sum_{i=1}^K {\bf s}_i({\pmb{\theta}})||^2.\\
{\pmb{\theta}}\quad\quad\quad\quad\quad\quad\quad\label{s5}
\end{array}
\end{equation}
It is equivalent to minimize the distance between the observed visibilities in ${{\bf{y}}}$, and the predicted interferometer response as a superposition of $K$ non-linear functions ${\bf s}_i(\pmb{\theta})$ for $i\in \{1,2,\ldots,K\}$. However, solving Eq. (\ref{s5}) suffers the same set of problems faced by any non-linear optimization problem, such as convergence to a local minimum, having a slow speed of convergence and significant computational cost.\\
There are various gradient-based optimization algorithms for estimating $\widehat{{\pmb{\theta}}}$ at Eq. (\ref{s5}). The iterative LM algorithm is one of the most robust gradient-based optimization techniques in the sense that most of the time, given suitable initial suggestion ${\pmb{\theta}}^0$, it converges to a global optimum. Considering $\phi({\pmb{\theta}})=||{\bf y}-\sum_{i=1}^K {\bf s}_i({\pmb{\theta}})||^2$ as the cost function, the estimation of ${\pmb{\theta}}$ at the $k+1$-th iteration of the algorithm will be obtained by
\begin{equation}
{\pmb{\theta}}^{k+1}={\pmb{\theta}}^k-({\pmb{\bigtriangledown}}_{\pmb{\theta}} {\pmb{\bigtriangledown}}^T_{\pmb{\theta}} \phi({\pmb{\theta}})+\lambda {\bf H})^{-1}{\pmb{\bigtriangledown}}_{\pmb{\theta}} \phi({\pmb{\theta}})|_{{\pmb{\theta}}^k}.\label{s6}
\end{equation}

In Eq. (\ref{s6}), ${\pmb{\bigtriangledown}}_{\pmb{\theta}}$ is the gradient with respect to ${\pmb{\theta}}$, and $\lambda$ is the damping factor which should be adjusted at each iteration \citep{Dampingterm}. The matrix ${\bf H}=\mbox{diag}({\pmb{\bigtriangledown}}_{\pmb{\theta}} {\pmb{\bigtriangledown}}^T_{\pmb{\theta}} \phi({\pmb{\theta}}))$  is the diagonal of the Hessian matrix. \\
The EM algorithm, and in particular the SAGE algorithm improve the accuracy and computational cost compared with the LS calibration. Since they break the ML estimation problem into smaller problems, the computational cost will be decreased by an order of magnitude and the rate of convergence is substantially increased.
 
\subsection{The EM calibration algorithm}\label{EM}

In order to apply the EM algorithm \citep{M.2} in calibration, we first need to extract a complete data space ${\bf x}$ from the observed data ${\bf y}$. Similar to \citet{S.2}, we consider the complete data space as ${\bf x}=[{\bf x}_1^T\ {\bf x}_2^T\ldots{\bf x}_K^T]^T$ in which ${\bf x}_i$ has the definition
\begin{equation}
{\bf x}_i\equiv{\bf s}_i({\pmb{\theta}}_i)+{\bf n}_i,\ \ for\ i\in\{1,2,\ldots K\}.\label{s7}
\end{equation}

In fact, Eq. (\ref{s7}) assumes that the contribution of the $i$-th source in the observation depends only on a subset of parameters, ${\pmb{\theta}}_i$. So, we partition the unknowns over the parameter vector ${\pmb{\theta}}$ corresponding to all the $K$ sources in the sky as ${\pmb{\theta}}=[{\pmb{\theta}}^T_1\ {\pmb{\theta}}^T_2\ldots{\pmb{\theta}}^T_K]^T$. This partitioning is justifiable as each source is at a unique direction on the sky, and for each \san{antenna}, the signal path of all sources is the same. This is the case for our initial assumption where the sources are separated sufficiently. Also, the total noise is arbitrary decomposed into $K$ components, ${\bf n}_i$ for $i\in\{1,2,\ldots,K\}$, such that
\begin{equation}
{\bf n}=\sum_{i=1}^K {\bf n}_i.\label{s8}
\end{equation}  
As the most convenient assumption, we let the noise components ${\bf n}_i$s to follow statistically independent zero mean Gaussian distributions with the covariance matrix
\begin{equation}
\operatorname{E}\{{\bf n}_i{\bf n}_j^H\}=\beta_i\delta_{ij}{\pmb{\Pi}},\label{bbb1}
\end{equation} 
where
\begin{equation}
\beta_i\in[0,1],\ \ \mbox{for}\ i\in\{1,2,\ldots,K\},\ \ \sum_{i=1}^K \beta_i=1.
\end{equation}
In principle, we can associate stronger sources with higher noise, hence higher $\beta_i$s, and vice-versa.\\
Combining  Eq. (\ref{s4}) and Eq. (\ref{s7}), the observed data ${\bf y}$ will be derived from
\begin{equation}
{\bf y}=\sum_{i=1}^K {\bf x}_i.\label{s77}
\end{equation}
Therefore, for the given complete data space ${\bf x}$ we have
\begin{equation}
{\bf y}=[{\bf I}\ {\bf I}\ldots{\bf I}]{\bf x}={\bf{Gx}},\label{s9}
\end{equation}
where ${\bf G}$ is a block matrix containing the identity matrix ${\bf I}$ for $K$ times.\\
Having the definitions of complete and observed data in hand, the EM algorithm can be used to estimate the ML of the parameter vector ${\pmb{\theta}}$. Applying the EM method for the new form of the measurement equation, Eq. (\ref{s9}), the below Expectation (E) and Maximization (M) steps are developed at the $k+1$-th iteration for $i\in\{1,2,\ldots,K\}$.\\

{\em E Step}: Calculating the conditional mean $\widehat{{\bf x}}_i^k=\operatorname{E}\{{\bf x}_i|{\bf y},{\pmb{\theta}}^k\}$. Considering the fact that ${\bf x}$ and ${\bf y}$ are jointly Gaussian we get 
\begin{equation}
\widehat{{\bf x}}_i^k={\bf s}_i({\pmb{\theta}}_i^k)+\beta_i({\bf y}-\sum_{l=1}^K {\bf s}_l({\pmb{\theta}}_l^k)).\label{s10}
\end{equation}  

{\em M Step}: Finding ${\pmb{\theta}}_i^{k+1}$ such that minimizes the cost function $\phi_i({\pmb{\theta}}_i)=||[\widehat{{\bf x}}_i^k-{\bf s}_i({\pmb{\theta}}_i)](\beta_i {\pmb{\Pi}})^{-\frac12}||^2$ with respect to ${\pmb{\theta}}_i$. This is also a non-linear optimization problem where the LM technique can be applied. The result is given by:
\begin{equation}
{\pmb{\theta}}_i^{k+1}={\pmb{\theta}}_i^k-({\pmb{\bigtriangledown}}_{{\pmb{\theta}}_i} {\pmb{\bigtriangledown}}^T_{{\pmb{\theta}}_i} \phi_i({\pmb{\theta}}_i)+\lambda {\bf H}_i)^{-1}{\pmb{\bigtriangledown}}_{{\pmb{\theta}}_i} \phi_i({\pmb{\theta}}_i)|_{{\pmb{\theta}}_i^k},\label{s11}
\end{equation}
where ${\bf H}_i=\mbox{diag}({\pmb{\bigtriangledown}}_{{\pmb{\theta}}_i} {\pmb{\bigtriangledown}}^T_{{\pmb{\theta}}_i}\phi_i({\pmb{\theta}}_i))$.\\

We repeat the above two steps starting from iteration $k=1$ until convergence or an upper limit which has been reached. Since at each iteration, $i$ goes from $1$ to $K$, the solutions of each source will be updated. In Appendix \ref{appen1}, we derive the EM algorithm for the problem in details and the results given above.
 
\subsection{The SAGE calibration algorithm}\label{SAGE}

SAGE algorithm \citep{J.A.1} performs better than the EM algorithm and has a higher speed of convergence and solution accuracy. The major difference between these two approaches is in the way of assigning the noise to the complete data space.

Similar to applying the classical EM algorithm, the first step in the SAGE algorithm is to find a complete data space relating the observations to the unknown parameters. For this purpose, consider the set of all indices related to all the $K$ sources as 
\begin{equation}
P=\{1,2,\ldots,K\}.\label{n12}
\end{equation}
Then, define index sets $W_i$ such that 
\begin{equation}
\varnothing\neq W_i\subsetneq P,\label{n13}
\end{equation}
where for all  $a,b\in W_i$, we have $a\neq b$, and the sources corresponding to the indexes $a$ and $b$, source $a$ and source $b$, have a small angular distance (they are near to each other in the sky) and subsequently they share some elements of the parameter vector ${\pmb{\theta}}$. We have
\begin{equation}
W_i\cap W_j=\varnothing,\ \ \mbox{for}\ i\neq j. \label{n14}
\end{equation}
Let us assume that we have $m$ such index sets. Thus,
\begin{equation}
m\le K,\ \ P=\bigcup_{i=1}^m W_i.\label{n15}
\end{equation}
For each $i\in\{1,2,\ldots,m\}$, we define a new parameter vector ${\pmb{\theta}}_{W_i}$ consisting of all the elements in the parameter vector ${\pmb{\theta}}$, which are affected by the sources with indexes in $W_i$. In other words, we provide the possibility to have elements in these new parameter vectors which are shared by more than one source.

Now, we make a new partitioning over the parameter vector ${\pmb{\theta}}$ as
\begin{equation}
{\pmb{\theta}}=[{\pmb{\theta}}_{W_1}^T\ {\pmb{\theta}}_{W_2}^T\ \ldots\ {\pmb{\theta}}_{W_m}^T]^T.\label{n16}
\end{equation}
 Similar to \citet{J.A.1}, we define the hidden data space ${\bf x}_{W_i}$ as
\begin{equation}
{\bf x}_{W_i}=\sum_{l\in {W_i}}{\bf s}_l({\pmb{\theta}}_{W_i})+{\bf n},\label{s12}
\end{equation}
 selecting the index set $W_i\in\{W_1,W_2,\ldots,W_m\}$ which preferably consists of the indices of the brightest sources. Note that in Eq. (\ref{s12}) all the noise has been associated to the sources with indices in ${W_i}$. This is the main difference between the SAGE and the classical EM algorithm. Using Eq. (\ref{s12}), the measurement equation can be written as
\begin{equation}
{\bf y}={\bf x}_{W_i}+\sum_{\substack{j=1\\j\neq i}}^{m}\sum_{l\in {W_j}}{\bf s}_l({\pmb{\theta}}_{W_j}).\label{s13}
\end{equation}
By applying the EM algorithm to this new form of the measurement equation, we arrive at the following steps for the $k+1$-th iteration of the SAGE approach:

{\em {SAGE E Step}}: Computing the conditional mean $\widehat{{\bf x}}_{W_i}^k=\operatorname{E}\{{\bf x}_{W_i}|{\bf y},{\pmb{\theta}}^k\}$. Since  ${\bf x}_{W_i}$ and ${\bf y}$ are also jointly Gaussian, we get
\begin{eqnarray}
\widehat{{\bf x}}_{W_i}^k&=&\sum_{l\in {W_i}}{\bf s}_l({\pmb{\theta}}_{W_i}^k)+({\bf y}-\sum_{j=1}^{m}\sum_{l\in {W_j}}{\bf s}_l({\pmb{\theta}}_{W_j}^k))\nonumber\\
&=&{\bf y}-\sum_{\substack{j=1\\j\neq i}}^{m}\sum_{l\in {W_j}}{\bf s}_l({\pmb{\theta}}_{W_j}^k).\label{s14}
\end{eqnarray} 

{\em {SAGE M Step}}: Finding ${\pmb{\theta}}_{W_i}^{k+1}$ which is minimizing the cost function $\phi_{W_i}({\pmb{\theta}}_{W_i})=||[\widehat{{\bf x}}_{W_i}^k-\sum_{l\in {W_i}}{\bf s}_l({\pmb{\theta}}_{W_i})]({\pmb{\Pi}})^{-\frac12}||^2$ with respect to ${\pmb{\theta}}_{W_i}$. The result is similar to Eq. (\ref{s11}). As before, we iterate from $k=1$ to an upper limit. At each iteration, we change the index set  $W_i$ within $\{W_1,W_2,\ldots,W_m\}$ to update all or some sources.\\

A special case of the SAGE algorithm was presented by \citet{S.2} if we consider $W_i=\{i\}$ for all $i\in\{1,2,\ldots,K\}$. In fact, we apply the same partitioning over the unknown parameter ${\pmb{\theta}}$ which is used for the classical EM algorithm,  ${\pmb{\theta}}=[{\pmb{\theta}}^T_1\ {\pmb{\theta}}^T_2\ldots{\pmb{\theta}}^T_K]^T$. By choosing the index $i$, where the source $i$ is preferably the brightest source in the sky, the hidden data space will be defined by
\begin{equation}
{\bf x}_{i}={\bf s}_i({\pmb{\theta}}_i)+{\bf n}.\label{n17}
\end{equation}
Eq. (\ref{n17}) gives us the definition of the observed data as
\begin{equation}
{\bf y}={\bf x}_i+\sum_{\substack{l=1\\l\neq i}}^K{\bf s}_l({\pmb{\theta}}_l),\label{n18}
\end{equation}
and subsequently, applying the EM on the measurement equation, the $k+1$-th iteration of the SAGE technique will be as below:

{\em {SAGE E Step}}: Conditional mean $\widehat{{\bf x}}_i^k=\operatorname{E}\{{\bf x}_i|{\bf y},{\pmb{\theta}}^k\}$ is derived from
\begin{equation}
\widehat{{\bf x}}_i^k={\bf s}_i({\pmb{\theta}}_i^k)+({\bf y}-\sum_{l=1}^K{\bf s}_l({\pmb{\theta}}_l^k))={\bf y}-\sum_{\substack{l=1\\l\neq i}}^K{\bf s}_l({\pmb{\theta}}_l^k).\label{n19}
\end{equation} 

{\em {SAGE M Step}}: ${\pmb{\theta}}_i^{k+1}$ is given by minimizing the cost function $\phi_i({\pmb{\theta}}_i)=||[\widehat{{\bf x}}_i^k-{\bf s}_i({\pmb{\theta}}_i)]({\pmb{\Pi}})^{-\frac12}||^2$.\\\\
In Appendix \ref{appen2}, we present the complete calculation process of applying the SAGE algorithm to the calibration problem.

\subsection{Computational Cost}\label{Computation}

At each iteration of the LS calibration scheme via the LM algorithm, the non-linear system presented by Eq. (\ref{s6}) should be solved which is of order $(KN)$. Therefore, ignoring the cost of calculations for the inverse part in this equation, the computational cost will be $\mathcal{O}((KN)^2)$. Furthermore, for radio synthesis arrays such as LOFAR and SKA, computing the matrix inverse in  Eq. (\ref{s6}) is very costly since the number of measured data is becoming very large. While, at each iteration of the EM algorithm, we should  solve Eq. (\ref{s11}) $K$ times and subsequently  the computational cost of the EM calibration algorithm will be $K\mathcal{O}(N^2)$, which is still much cheaper compared with the LS calibration approach. Thus, the  EM as well as the SAGE calibration techniques are superior to the LS one in terms of computational cost.\\
\san{Note that the LM optimization technique is employed for all the LS, EM, and the SAGE calibration algorithms. Thus, its corresponding inversion computation is shared in all the methods. However, in the EM and SAGE algorithms the size of the matrix that is inverted is smaller compared to that of the LS algorithm because of the partitioning procedure of the parameters. Given the fact that the computational complexity of the matrix inversion scales a number of its elements to the third power, it is evident that inverting few smaller matrices as in the EM and SAGE cases is faster than inverting a single large matrix, which is the case for the LS algorithm.} 

\section{NOISE IN SOLUTIONS}\label{Solvernoise}
To compare the accuracy of the SAGE calibration scheme to the LS one, we statistically analyze the solver noise. The lower the solver noise in a calibration method, the smaller the errors in calibrated solutions provided by the calibration algorithm itself. Consequently, the accuracy of the method is higher.

In order to do a proper statistical analysis of the calibration algorithm's solver noise, we
 make the assumption that their solutions are linear combinations
of a deterministic trend and noise. This noise can have many origins. In the ideal case,
it is introduced by the primary noise sources (thermal noise at the receiver, the sky noise, radio interference, etc.) and by variations of the instrumental and propagation properties. However, in reality the solver noise, which we are most concerned about and is introduced by the calibration method itself, is also added to thoese sources.

 The goal in this section is to quantify the level of the solver noise for the different calibration algorithms, based on the  evaluation of the statistical interaction between their solutions.

\subsection{Statistical similarity}\label{SS}
In order to compare the level of solver noise for the different calibration methods, we assume that the true values of the solutions from different directions at the same \san{antenna} are statistically uncorrelated. Therefore, any correlation between the calibrated solutions for different directions is caused by the corresponding calibration technique itself. In reality, there are also correlations that originate from the system noise in the solutions, but this can be ignored when we compare the solutions of a fixed measured data obtained by different calibration methods. Therefore, a high solver noise in a calibration scheme causes strongly correlated solutions for any number of directions at one given \san{antenna} (or maybe even more). To detect the statistical similarity between the gain solutions we proceed as follows:

Once we estimate the parameter vector ${\pmb{\theta}}$, we obtain the Jones matrices ${{\bf{J}}}_{qs}({\pmb{\theta}})$ for different \san{antennas} $q\in\{1,2,\ldots,N\}$ and different directions $s\in\{1,2,\ldots,K\}$. Let us consider the matrices to be diagonal as \begin{equation}
\textbf{J}_{qs}=\left[\begin{array}{cc}
J_{11,q}&0\\
0&J_{22,q}\end{array}\right]_s,\label{exam}
\end{equation}
where $J_{11,q}$ and $J_{22,q}$ are complex values. 
We treat each gain solution from the direction $s$ of antenna $q$
as a random vector ${\pmb{\theta}}_{qs}$ defined by 
\begin{equation}
\pmb{\theta}_{qs}=[\mathfrak{Re}(J_{11,q})\
\mathfrak{Im}(J_{11,q})\ \mathfrak{Re}(J_{22,q})\ \mathfrak{Im}(J_{22,q})]_s^T.\label{s20}
\end{equation}
Now, we can investigate the statistical similarity between the gain solutions utilizing Kullback-Leibler Divergence (KLD) and Likelihood-Ratio Test (LRT). In general, both KLD and LRT compare the efficiency of fitting two different statistical models to a fixed set of measurements. Utilizing these methods on the random vectors defined by Eq. (\ref{s20}), we obtain their statistical similarity in two different interpretations. The higher these similarities, the higher the interaction between the solutions, as well as the solver noise. 

\subsection{Kullback-Leibler Divergence (KLD)}

An efficient way to quantify the statistical similarity between the solutions is to use KLD. 

The relative entropy, defined as the KLD, for each couple of Probability Density Functions (PDFs) $f$ and $g$ of solutions $\pmb{\theta}_{qk}$ and $\pmb{\theta}_{ql}$, respectively, is defined as
\begin{equation}
\mbox{KLD}(f,g)\equiv\sum_{\pmb{\theta}_{qk}}^{}f(\pmb{\theta}_{qk})\mbox{log}\frac{f(\pmb{\theta}_{qk})}{g(\pmb{\theta}_{qk})},\label{1}
\end{equation}
where the solutions are corresponding to the given antenna $q$ at the two different directions $k$ and $l$.\\
The KLD is a measure of information ``divergence'' between two different PDFs for the same random variable. Larger values of $\mbox{KLD}(f,g)$ are interpreted as less interaction
between the solutions, and subsequently, as less solver noise. We use a Monte-Carlo method to evaluate Eq. (\ref{1}). 
\subsubsection{Density estimation}

To calculate the value of KLD we need to estimate the PDFs of the solutions.

We define the PDF of the random
vector $\pmb{\theta}_{qs}$, $f(\pmb{\theta}_{qs};{\pmb{\beta}})$, as a
mixture of $L$ isotropic (scalar variance) Gaussian PDFs. \san{The assumption of mixture modeling is based on the fact that the solutions are affected by parameters that belong to different underlying statistical populations and due to the Central Limit Theorem it is reasonable to assume that the distributions of those populations converge to Gaussian distributions.}. Therefore, it can be written as
\begin{equation}
f(\pmb{\theta}_{qs},{\pmb{\beta}})=\sum_{l=1}^{L}p_lg(\pmb{\theta}_{qs};\textbf{m}_l,\sigma_l),
\label{2}
\end{equation}
where $g(\pmb{\theta}_{qs};\textbf{m}_l,\sigma_l)$ is the PDF of
a four dimensional Gaussian distribution with mean $\textbf{m}_l$
and variance $\sigma^2_l \textbf{I}$ given by\\
\begin{equation}
g(\pmb{\theta}_{qs};\textbf{m}_l,\sigma_l)=\frac{1}{(\sqrt{2\pi}\sigma_l)^4}exp\bigg(-\frac12\bigg(\frac{\|\pmb{\theta}_{qs}-\textbf{m}_l\|}{\sigma_l}\bigg)^2\bigg),
\end{equation}\\
and ${\pmb{\beta}}$ is the vector of the mixture model unknown
parameters\\
\begin{equation}
{\pmb{\beta}}=[p_1,\textbf{m}_1^T,\sigma_1,...,p_L,\textbf{m}_L^T,\sigma_L]^T,
\end{equation}\\
which are estimated by the EM algorithm \citep{citeulike:3334839}. 

\subsubsection{Akaike's Information Criterion for model order selection}
To find the optimum number of $L$ (the order of Gaussian mixture model in Eq. (\ref{2})) we use Aikake's Information Criterion (AIC, \citealt{H.1}). 

According to the definition of AIC, we select a $L$ such that it gives us the minimum value of AIC($L$) which is  defined as
\begin{equation}
\mbox{AIC}(L)=-2\mbox{L}(\widehat{{\pmb{\beta}}})+2k,
\end{equation}
where $\mbox{L}(.)$ is the log-likelihood function of $\pmb{\theta}$, $\widehat{{\pmb{\beta}}}$ is the maximum likelihood estimate of $\pmb{\beta}$, and $k$ is the number of parameters in the model presented by Eq. (\ref{2}). 

\subsection{Likelihood-Ratio Test (LRT)}

Another standard approach to investigate the statistical interaction between the solutions is the Likelihood-Ratio Test (LRT). Using this test, we can compare two models, which both can be fitted to our solutions.\\
Let us define for each antenna $q$, where $q\in\{1,2,...,N\}$, and each pair of directions like $k$ and $l$, where $k, l\in\{1,2,...,K\}$, a new random vector $\textbf{z}_{qkl}$ as
\begin{equation}
\textbf{z}_{qkl}=[{\pmb{\theta}}^T_{qk}\ {\pmb{\theta}}^T_{ql}]^T.\label{12}
\end{equation}
In fact, we are concatenating the solutions of the same antenna for two different directions together. Assume that $\textbf{z}_{qkl}$ is following a multivariate Gaussian distribution with  mean 
\begin{equation}
\textbf{m}=[\bar{\textbf{m}}({\pmb{\theta}}_{qk})^T\ \bar{\textbf{m}}({\pmb{\theta}}_{ql})^T]^T,\label{8}
\end{equation}
and variance 
\begin{equation}
\Sigma_\textbf{0}=\left[\begin{array}{cc}
s^2(\pmb{\theta}_{qk})& 0\\
0&s^2(\pmb{\theta}_{ql})\end{array}\right],\label{9}
\end{equation}
where $\bar{\textbf{m}}$ and $s^2$ are the sample mean and sample variance of the solutions respectively. The structure of the variance matrix $\Sigma_\textbf{0}$ tells us that the statistical correlation between the components of the random vector $\textbf{z}_{qkl}$, or between the solutions ${\pmb{\theta}}_{qk}$ and ${\pmb{\theta}}_{ql}$, is zero. This is exactly the desirable case in which the solver noise vanishes. Thus, we can consider this model as our null $H_0$ model defined by
 \begin{equation}
H_0:\ \textbf{z}_{qkl}\sim \mathcal{N}(\textbf{m},\Sigma_0).\label{13.1}
\end{equation}
To investigate the validity of the null model compared with the case in which there exist some correlation between the solutions due to the presence of solver noise, we define the alternative $H_1$ model as
\begin{equation}
H_1:\ \textbf{z}_{qkl}\sim \mathcal{N}(\textbf{m},\Sigma_1),\label{13.2}
\end{equation}
where the variance matrix $\Sigma_1$ is given by
\begin{equation}
\Sigma_1=\left[\begin{array}{cc}
s^2(\pmb{\theta}_{qk})& \textbf{Q}(\pmb{\theta}_{qk},\pmb{\theta}_{ql})\\[2mm]
\textbf{Q}(\pmb{\theta}_{qk},\pmb{\theta}_{ql})^T&s^2(\pmb{\theta}_{ql})\end{array}\right],
\end{equation}\\
and the $4\times 4$ matrix $\textbf{Q}(\pmb{\theta}_{qk},\pmb{\theta}_{ql})$ denotes the sample covariance of the solutions.\\
Using the above models, the Likelihood-Ratio is defined as
\begin{equation}
\Lambda=-2\mbox{ln}\bigg(\frac{\mbox{Likelihood for null model}}{\mbox{Likelihood for alternative model}}\bigg)\label{lrt1},
\end{equation}
which has a $\chi^2$ distribution with 16 degrees of freedom. As $\Lambda$ becomes smaller, the null model, in which the statistical correlation as well as the statistical similarity between the solutions is zero, becomes 
 more acceptable compared to the alternative one. Therefore, the smallest the  $\Lambda$, the less the solver noise and vice-versa.\\

Note that the test result is reliable only when a large number of sample solutions is in hand. In this case, because of the Central Limit Theorem, the distribution of solutions tends to be a multivariate LS distribution, which is assumed initially by the test.

\section{ILLUSTRATIVE EXAMPLES}\label{example}
\subsection{Simulated observation}

First we use a simulated observation to compare the efficiency of the SAGE calibration algorithm with the LS algorithm. Utilizing simulations instead of real observations has the advantage of having the true underlying gains available, which is a luxury not available with real data. Therefore, assessing the convergence of the calibration techniques as well as comparing the accuracy between different calibrated solutions is much more objective.  

\begin{figure*}
\centering
\hspace{1cm}
\includegraphics*[width=9.5cm,height=9.5cm, viewport=400 1 1300 850]{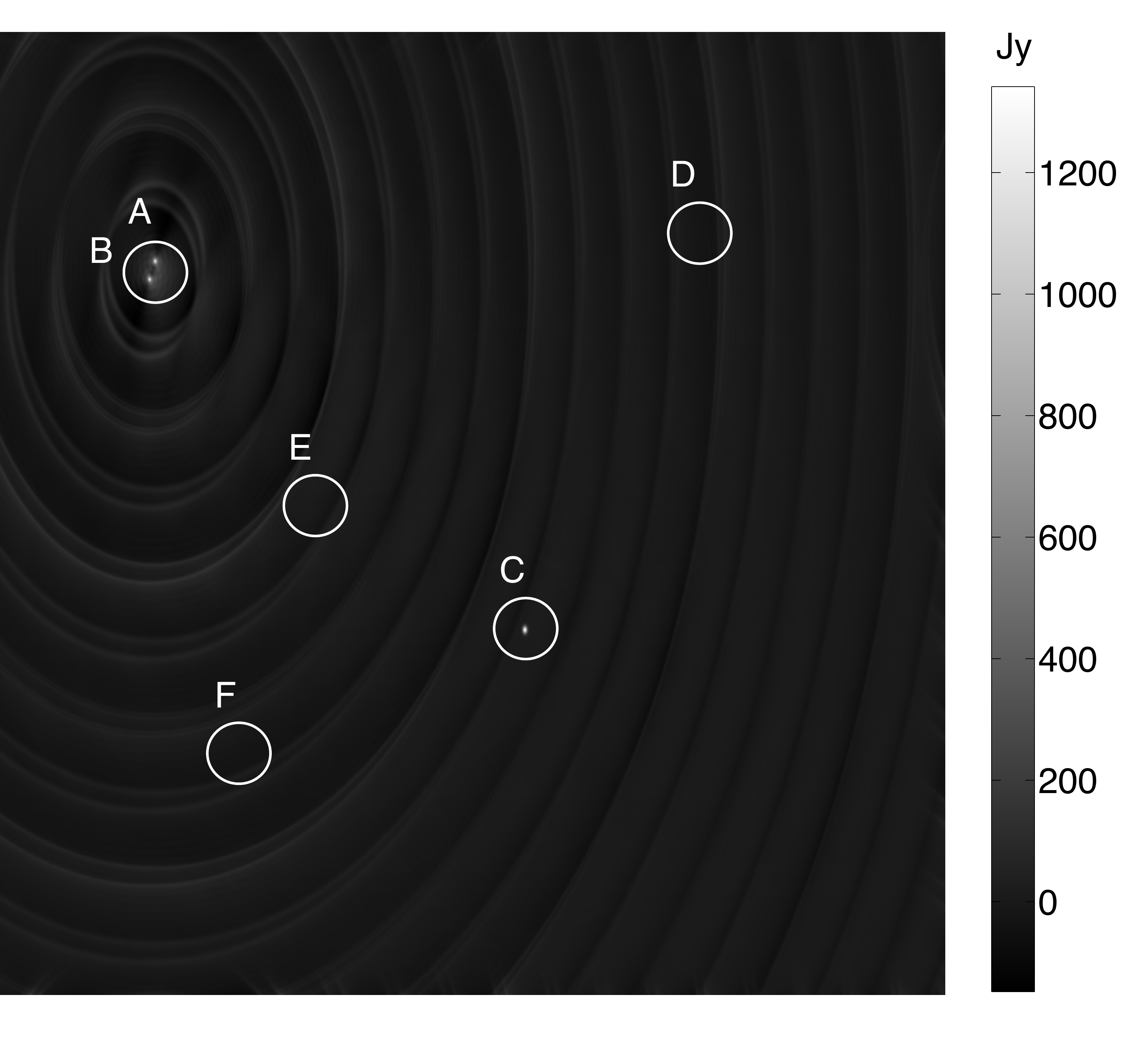}
\caption{\emph{Single channel simulated observation of three bright sources, A, B, and C, and three weak sources, D, E, and F. The intensity of the bright sources are 2950, 2900, and 2700 Jy and of the weak sources are 4, 3.5, and 3 Jy, respectively. The image size is 8 by 8 degrees at 355 MHz. There are no gain errors and noise in the simulation.  }}
\label{fig:first}
\end{figure*}

\begin{figure*}
\centering
\hspace{1cm}
\includegraphics*[width=9.5cm,height=9.5cm, viewport=400 1 1300 850]{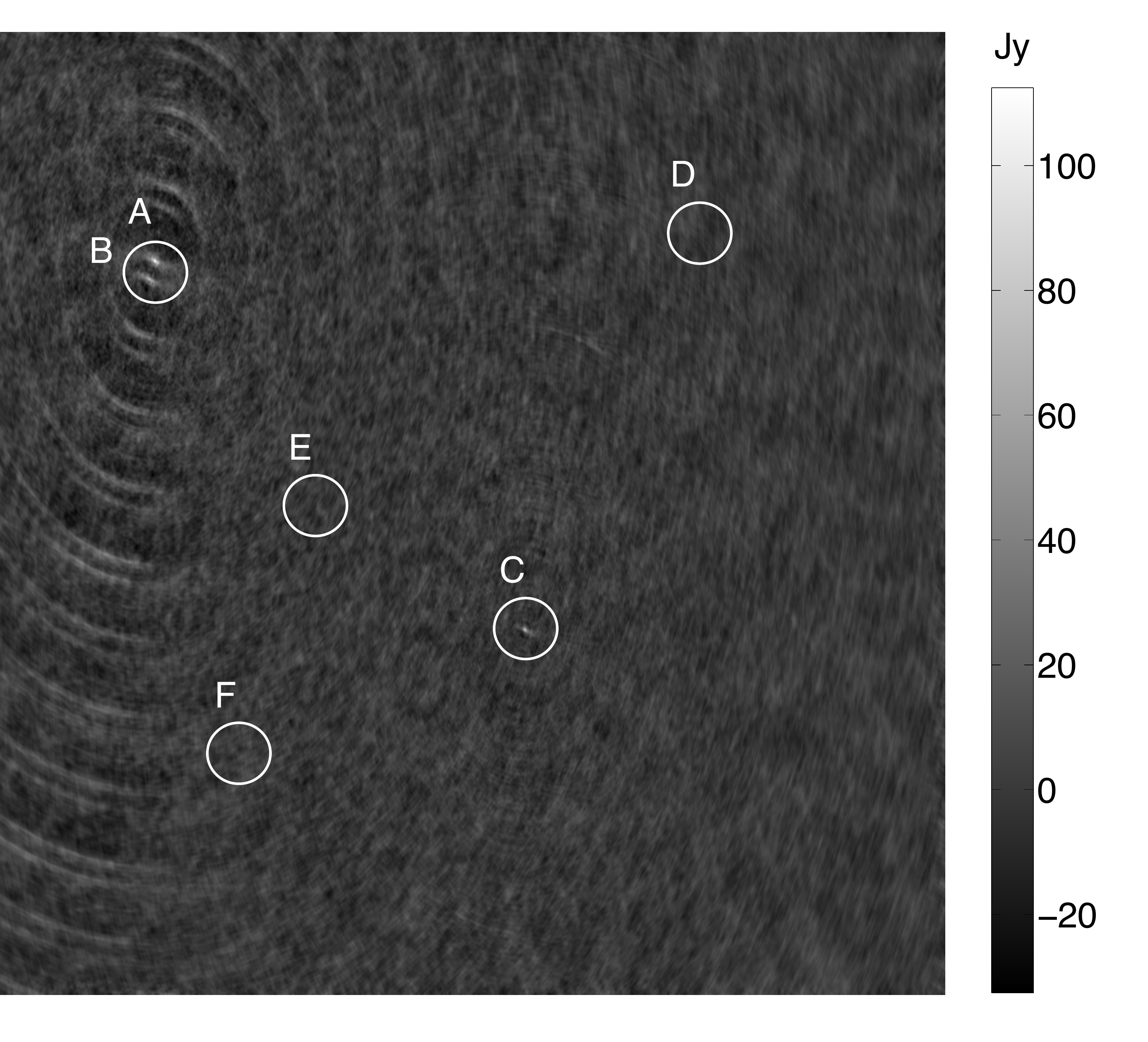}
\caption{\emph{Simulated observation with added gain errors. The errors are complex numbers having norms and phases as multiplications of random numbers with various linear combinations of $sin$ and $cos$ functions. The gradients of the errors increase as a function of time and the phases are in different negative and positive slopes. }}
\label{fig:gainy}
\end{figure*}

\begin{figure*}
\centering
\hspace{1cm}
\includegraphics*[width=9.5cm,height=9.5cm, viewport=400 1 1300 850]{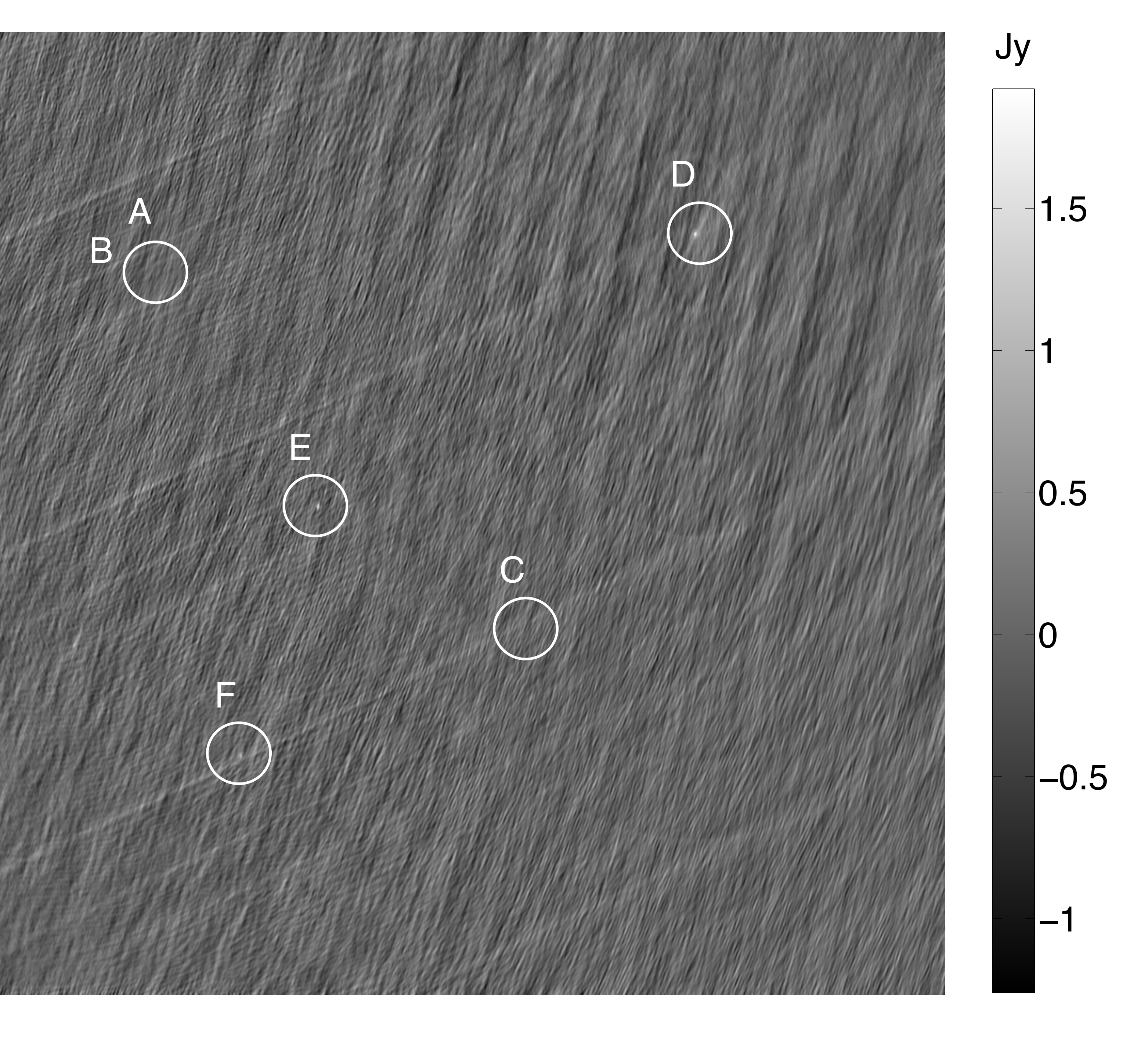}
\caption{\emph{Residual image of the SAGE calibration using nine iterations. The image is only calibrated for the bright sources A, B, and C, which are perfectly removed. The weak sources D, E, and F appear in the residuals, including source F which is the weakest source in the simulation.}}
\label{fig:SResid}
\end{figure*}

\begin{figure*}
\centering
\hspace{1cm}
\includegraphics*[width=9.5cm,height=9.5cm, viewport=400 1 1300 850]{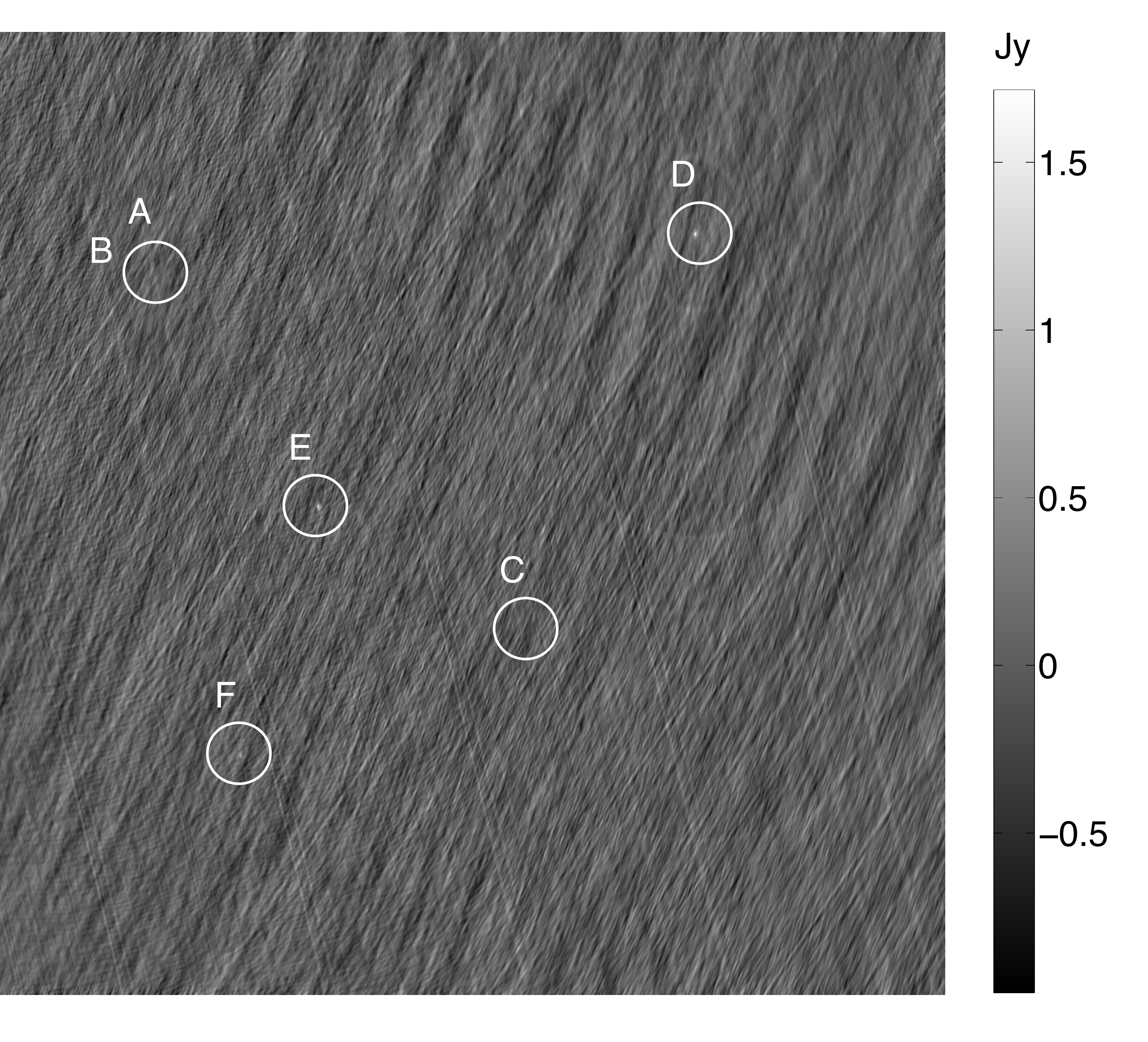}
\caption{\emph{Residual image of the LS calibration using nine iterations. The calibration is processed only for the strong sources A, B, and C, that are completely removed in the residuals. The two weak sources D, and E appear in the image, but the weakest source F is hardly visible.}}
\label{fig:NResid}
\end{figure*}

We consider a linear, East-West radio synthesis array that has 14 dipoles with dual polarization. We put three bright sources in our sky model named A, B, and C with intensities 2950, 2900, and 2700 Jy, and three other weak sources named D, E, and F with intensities 4, 3.5, and 3 Jy, respectively. The simulated single channel image at 355 MHz is shown in Fig. \ref{fig:first}. As we can see in Fig. \ref{fig:first}, the weak sources are not visible in the image. We also consider that there is no beam pattern, therefore the whole sky is being observed with uniform sensitivity.\\
As it is shown in Eq. (\ref{s2}), in the measured visibilities, the coherency of the sources are multiplied by the Jones matrices (gain errors). We consider the matrices to be diagonal. It means that the signal received at each dipole is not \san{affected} by the other one which is an ideal case. We produce gain errors in the norm and phase of the Jones matrices diagonal terms which are 1 and 0 initially. We generate the norm and the phase of the gains as multiplications of random numbers with different linear combinations of $sin$ and $cos$ functions, whose gradients increase with time. We also add another random term increasing as a function  of time, just to the phase errors, to provide the phases with positive and negative slopes. The simulated result is presented by Fig. \ref{fig:gainy}.\\
 Finally, we apply the SAGE and the LS calibration to solve only for the gains of the visible strong sources A, B, and C. The residuals of the SAGE and LS calibration using nine iterations are shown in Fig. \ref{fig:SResid} and Fig. \ref{fig:NResid}, respectively. 
As we can see in  Fig. \ref{fig:SResid} and Fig. \ref{fig:NResid}, The strong sources are completely removed and the three weak sources are visible in the residuals of the both algorithms which mean both are converging to real solutions. But, the weak sources intensities in the residuals of the SAGE calibration are closer to the absolute intensities in our sky model which shows the superiority of the SAGE calibration in terms of accuracy. This fact is shown more clearly by table \ref{table1} where the real intensities of the \san{weak} sources are \san{compared} with the calibrated ones. 
\begin{center}
    \begin{tabular}{ c|c c c}
     & {\bf{Table\ 5.1}}& & \\
    Source & D & E & F\\
    \hline\\[-3mm]
    Real intensity(Jy) & 4 & 3.5 & 3\\
    SAGE calibration & 3.8399 & 3.6695 & 2.3085\\
    LS calibration & 2.8327 & 2.6329 & 1.6081\\
    \hline
    
    \end{tabular}\label{table1}
\end{center}

\begin{figure}
\centering
\vspace{-0.0cm}
\hspace*{-7mm}
\includegraphics[width=9.5cm,height=6cm]{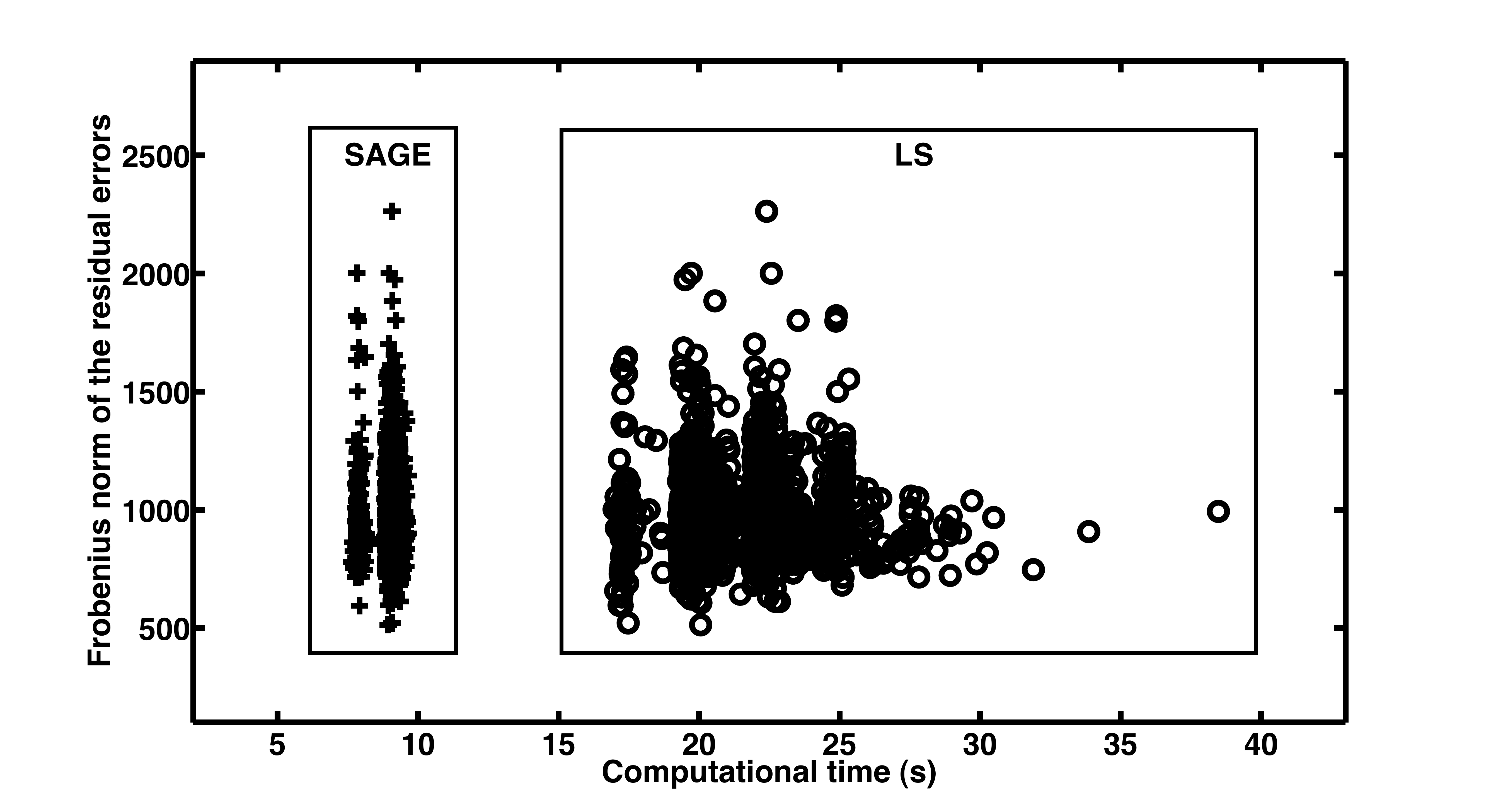}
\vspace{-0.5cm}
\caption{\emph{Comparison between the performance of the SAGE and the LS calibrations (nine number of iterations) in order of the accuracy of solutions and the speed of convergence. Different points correspond to  different snapshots made from the simulation. The speed of the SAGE calibration technique is higher than the LS calibration, while the norm of their residual errors are almost the same}}
\label{fig:SN9}
\end{figure}

Note that as this is a single channel simulated observation without any additive noise, the difference between the two method's residuals is slight. However, the importance of applying the SAGE instead of LS calibration is evident since the computational cost of the SAGE is much smaller, as it is discussed in section \ref{Computation}.  Fig. \ref{fig:SN9} shows the performance of the algorithms in terms of accuracy and speed of convergence. As we can see in  Figure \ref{fig:SN9}, the SAGE algorithm's speed of convergence is much higher than that of the standard LS calibration. 

\begin{figure}
\centering
\vspace{-0.0cm}
\hspace*{-5mm}
\includegraphics[width=9.5cm,height=6cm]{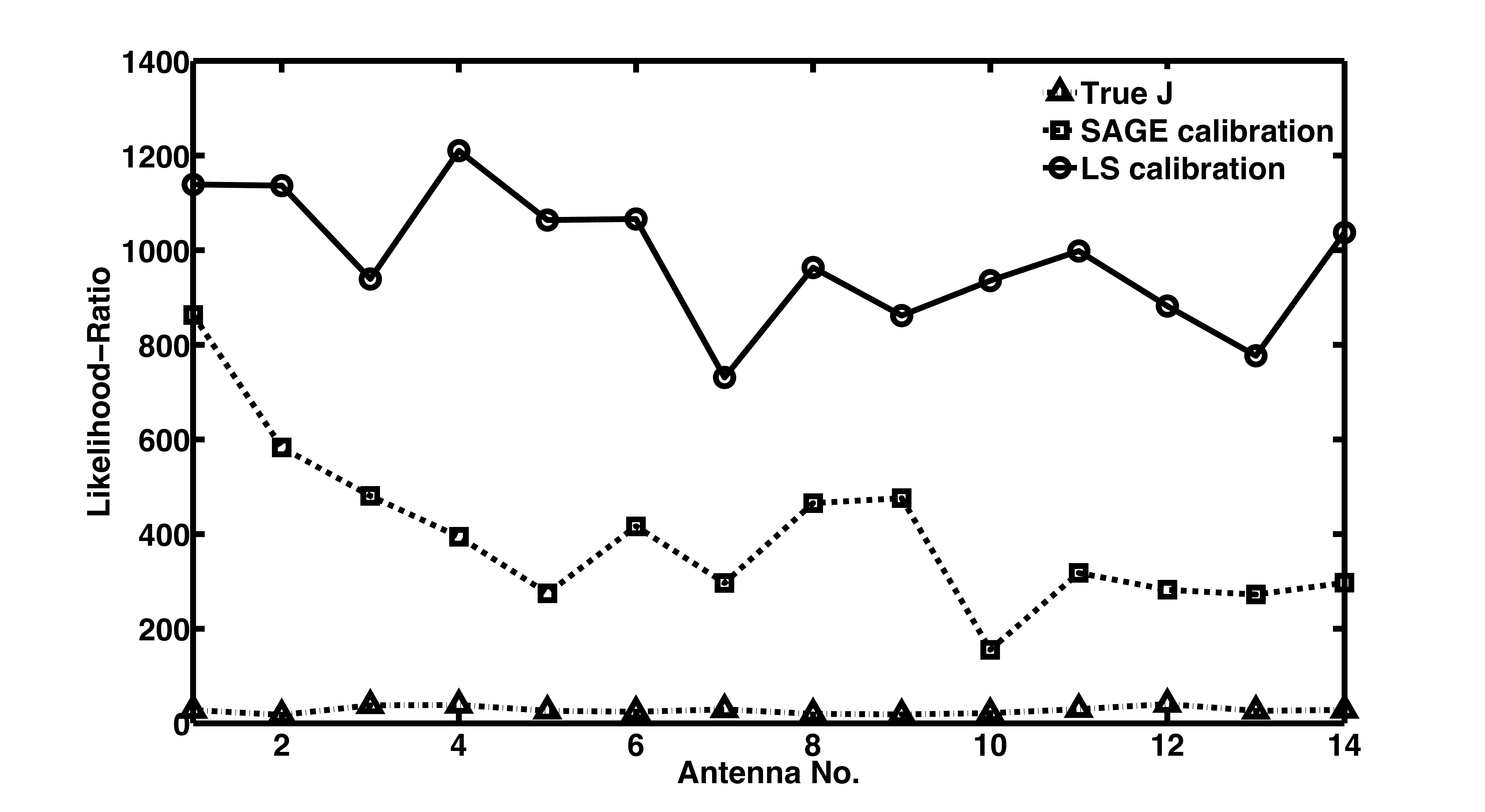}
\vspace{-0.5cm}
\caption{\emph{Averaged Likelihood-Ratio of direction dependent gains (True values and calibrated ones) for all two source combinations between the bright sources A, B, and C}}
\label{fig:LRTSim}
\end{figure}

\begin{figure}
\centering
\vspace{-0.0cm}
\hspace*{-5mm}
\includegraphics[width=9.5cm,height=6cm]{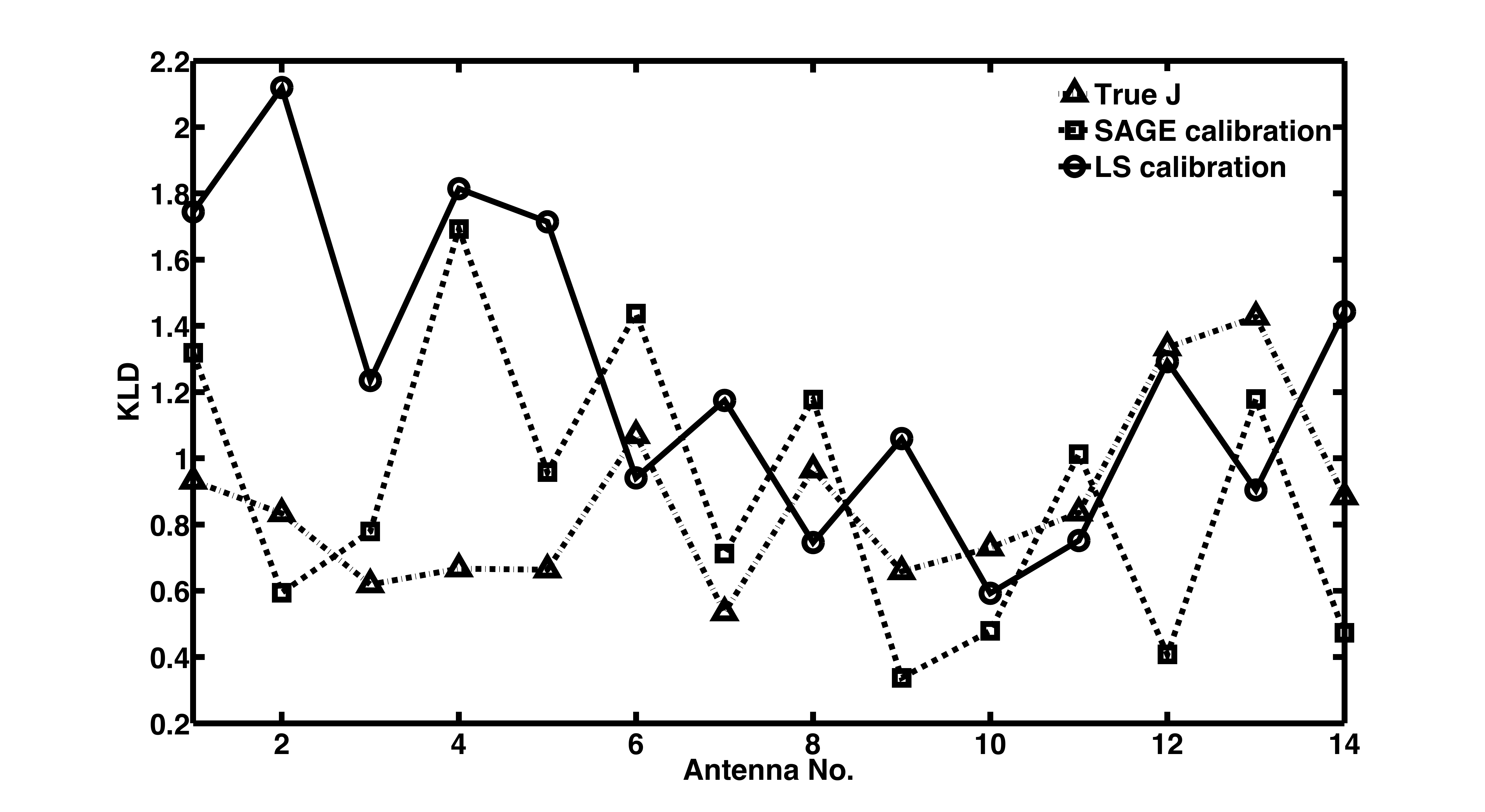}
\vspace{-0.5cm}
\caption{\emph{Averaged KLD of direction dependent gains (True values and calibrated ones) for all two source combinations between the bright sources A, B, and C}}
\label{fig:KLDSim}
\end{figure}

\san{In order to investigate the efficiency of the LRT and KLD approaches in revealing the statistical similarity of gain solutions, we applied both methods, using  Eq. \ref{1} and Eq. \ref{lrt1}, to the SAGE and the LS calibration solutions and compared their results with the LRT and KLD of the true Jones parameters (simulated gains). The comparison is shown by Fig. \ref{fig:LRTSim} and Fig. \ref{fig:KLDSim}.  Fig. \ref{fig:LRTSim} exhibits an outstanding performance of the LRT approach in which the direction dependent gain's lowest statistical similarity belongs to the real Jones values, that is almost zero. For SAGE calibration's solutions this similarity becomes higher, and in LS results it reaches to its highest level. This result demonstrates the superior accuracy of the SAGE calibration's solutions regardless of the residual images. But, the KLD results in Fig. \ref{fig:KLDSim} show the same level of statistical correlation in the calibrations' solutions and in the True Jones parameters. This is due to the fact that these results are calculated by Gaussian mixture models, which are fitted to the gains. This characteristic of the KLD approach (using fitted PDFs for gains distributions rather than the true PDFs) decreases the methods sensitivity in revealing the level of statistical similarity between different directions' gains, especially in our case where the simulated gains' correlations are low . However, we anticipate better performance of KLD method in real observations in which the higher gain errors plus the additive noise cause a higher solver noise.}

\subsection{Real observations}
\begin{figure*}
\centering
\vspace{-0.0cm}
\hspace*{-1.7cm}
\includegraphics[scale=0.6]{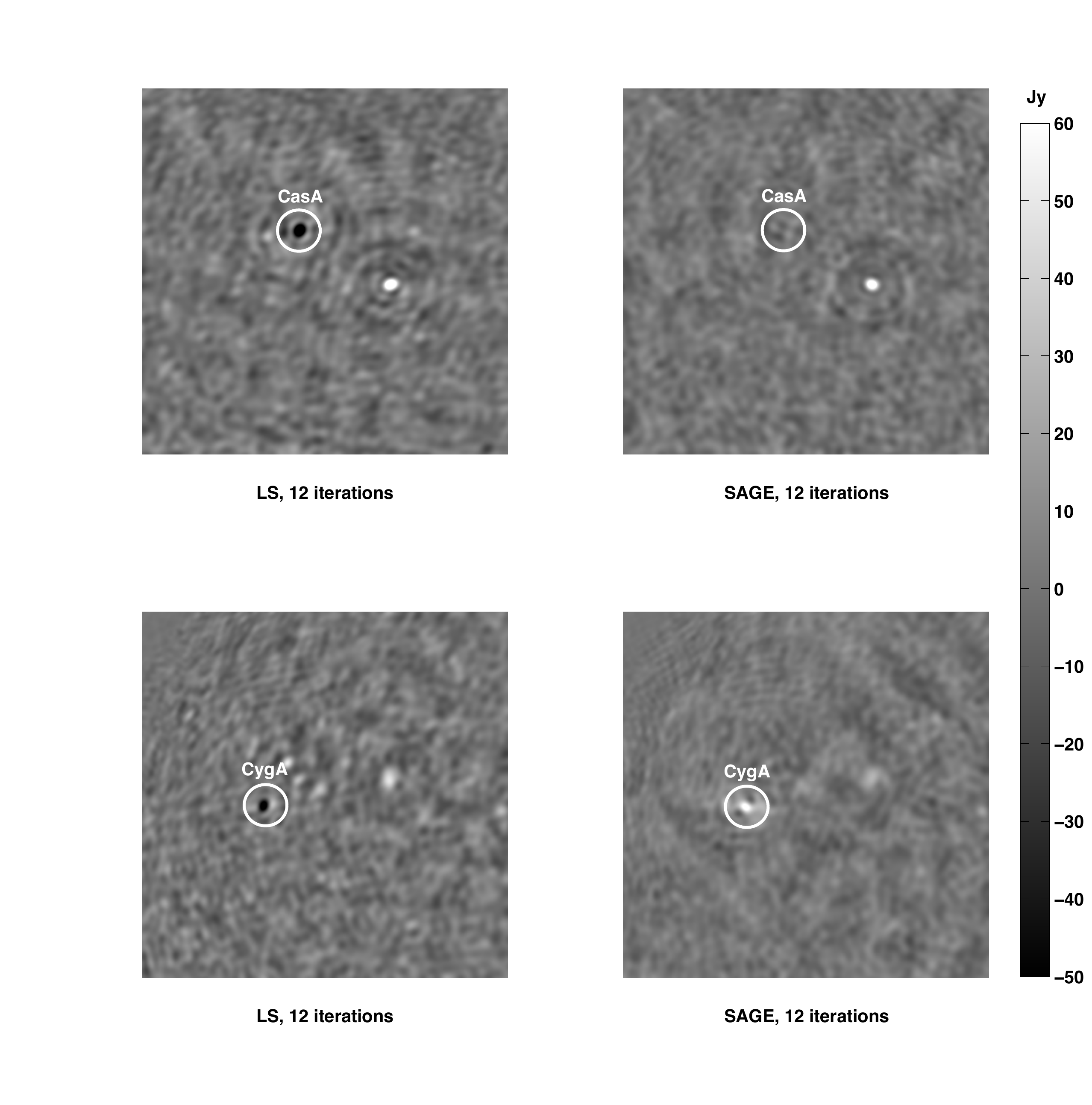}
\vspace{-0.5cm}
\caption{\emph{Residual images around CasA (top row) and CygA (bottom row) obtained by the LS calibration (left column) and the SAGE calibration (right column) using twelve iterations. CasA is over-subtracted in the residual of the LS calibration result, due to the inaccuracy in the estimation of the relevant direction-dependent gain, while in the SAGE calibration result it is removed almost perfectly. On the other hand, Both the SAGE and the LS calibrations present some problems around CygA since at some point of the integration time it goes very close to the horizon. Even in that case, the subtraction residual for the SAGE algorithm is 10 percent lower than for the LS method. }}
\label{fig:CasA&CygA}
\end{figure*}

\begin{figure}
\centering
\vspace{-0.0cm}
\hspace*{-5mm}
\includegraphics[width=9.5cm,height=6cm]{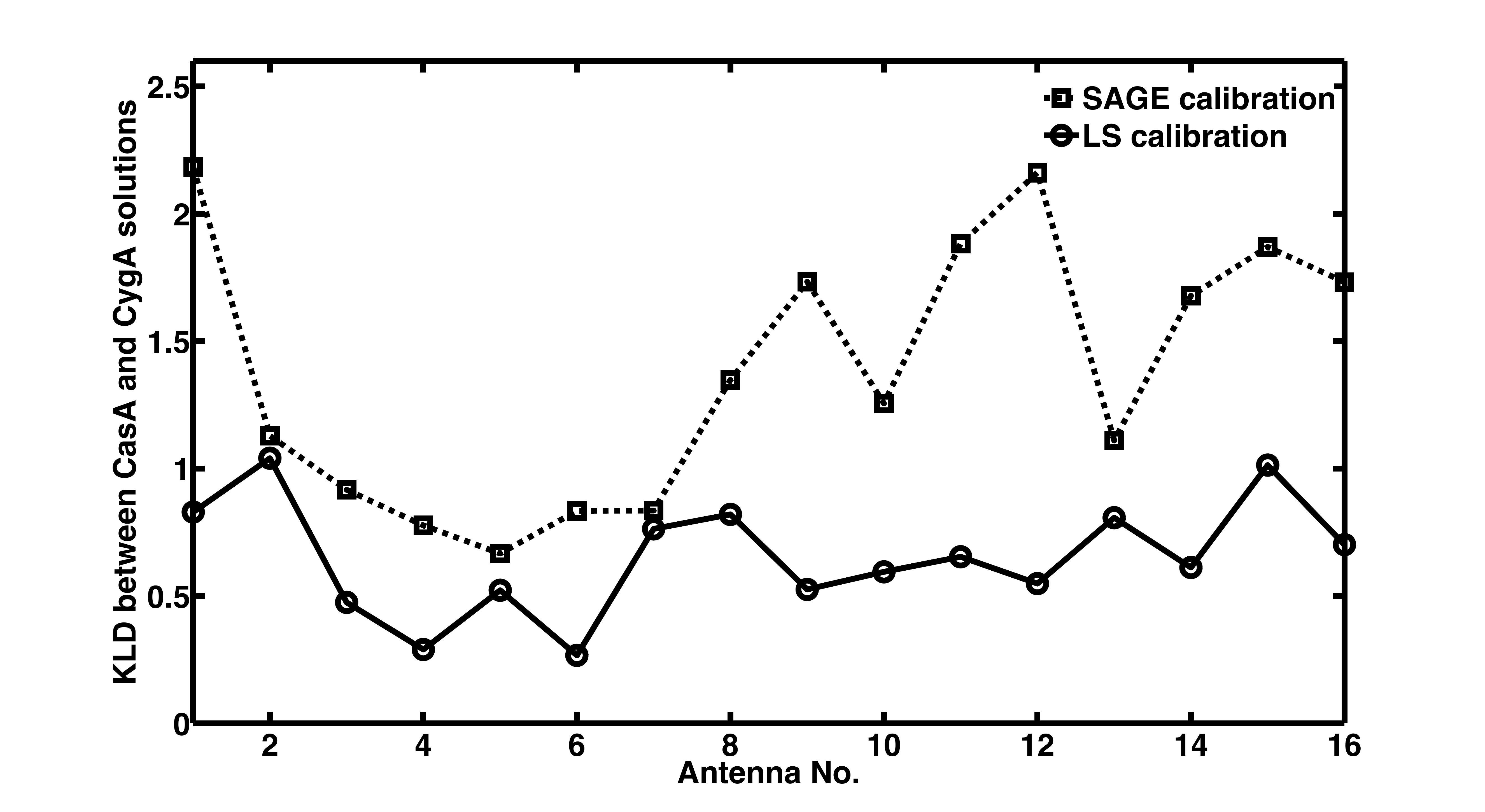}
\vspace{-0.5cm}
\caption{\emph{KLD of the gain solutions for CasA and CygA}}
\label{fig:KLD}
\end{figure}

\begin{figure}
\centering
\vspace{-0.0cm}
\hspace*{-5mm}
\includegraphics[width=9.5cm,height=6cm]{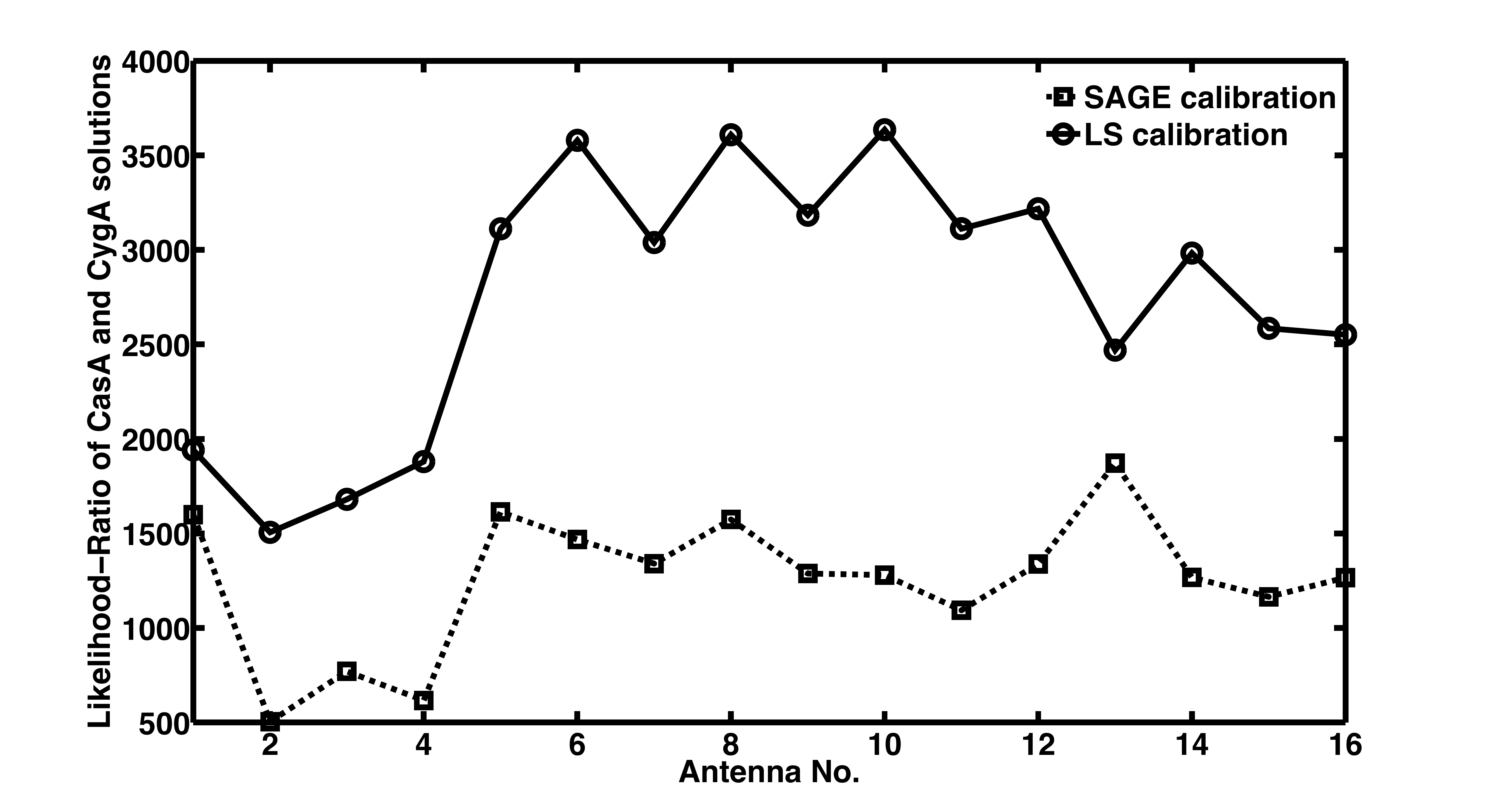}
\vspace{-0.5cm}
\caption{\emph{Likelihood-Ratio of the gain solutions for CasA and CygA}}
\label{fig:LRT}
\end{figure}

To illustrate the applicability of the KLD and LRT approaches in detecting solver noise in real observations calibrated solutions, we use the data from the example in \citet{S.2}. \citet{S.2} presents the calibration results using real data obtained during a 24 hours long LOFAR test core station (CS1) observation, via SAGE and LS calibration techniques. The one channel images around 3C 461 (Cassiopeia A, CasA) and 3C 401 (Cygnus A, CygA) at 50 MHz after applying these calibration methods using twelve iterations are shown in Fig. \ref{fig:CasA&CygA}. The result at Fig. \ref{fig:CasA&CygA} clearly verifies the superiority of the SAGE calibration scheme as it was mentioned in \citet{S.2} as well. 
We calculate the KLD using Eq. \ref{1} as well as Likelihood-Ratio  using Eq. \ref{lrt1} for the calibrated solutions of these two sources obtained by the mentioned calibration techniques. In the KLD approach, we fit a Gaussian mixture model, which its components have full rank covariance matrices, to the solutions. The KLD and LRT results are shown in Fig \ref{fig:KLD} and Fig \ref{fig:LRT}, respectively. As we can see in Fig \ref{fig:KLD}, the KLD of the solutions derived by the LS calibration is always lower than that of the SAGE calibration. It is also shown in Fig \ref{fig:LRT} that the LRT of the LS calibration's solutions is always higher than the SAGE calibration's. Therefore, the solver noise in the LS calibration results is measurably higher than in the solutions obtained by the SAGE calibration. This means that the accuracy of the SAGE calibration is always higher than that of the LS calibration, which is visible in Fig. \ref{fig:CasA&CygA} as well as the aforementioned images in \citet{S.2}. 

\section{Summary}
\label{sec:summary}
Since the new generation of radio synthesis arrays are producing a large amount of data with high sensitivity, it is of great interest to devise new calibration techniques in order to increase the accuracy of solutions with the highest possible speed of convergence.

In this paper, we presented the superior performance of the SAGE calibration scheme compared with the traditional LS calibration method. The superiority is in the sense that SAGE calibration has the highest accuracy, the fastest speed of convergence, and the cheapest computational cost. Since both the algorithms are estimating the ML of unknowns in different ways, it is possible that in some special cases ,such as having a very low initial noise in the measured visibilities, we don't have a specific difference between the accuracy of their solutions. While, even in this case, the SAGE calibration's faster speed of convergence and cheaper computational cost justify its application instead of the LS calibration. We compared the accuracy and the rate of convergence of the SAGE and the LS calibration in a simulated observation example. More accurate results in a much shorter time are obtained by the SAGE algorithm compared with the LS. The challenge in improving the performance of the SAGE calibration technique is to find the best way of partitioning over the unknown parameter space. This can highly affect both the accuracy and speed of convergence of the calibration process.

On the other hand, there always exists some estimation errors in the calibrated solutions. These errors are originated from the system noise (sky and instrumental) in the measurements, plus ``solver noise'' which is referred to errors produced by the calibration algorithm itself. The more accurate the calibrated solutions are, the less the amount of solver noise is. Based on this fact, the best calibration method is the one which provides us with  the minimum solver noise. KLD and LRT are utilized to reveal the level of solver noise in the solutions of different calibration schemes. We showed in \san{illustrative} examples that the LRT algorithm produces a very  promising result. The KLD method is rather inconclusive according to the initial assumption for the PDFs fitted to the solutions. We assumed that the distribution of the solutions is a mixture of Gaussian distributions. However, in reality, the solutions may follow a different distribution and subsequently the KLD result \san{may not have the same efficiency as the LRT's}. Therefore, initially we should find the proper distribution which is appropriate for the solutions in order to calculate the KLD. 

The main direction of future work should be  to investigate the application of the proposed calibration technique to real data obtained by LOFAR. Since LOFAR is observing the whole sky, the number of radio sources in the sky model will be very large. Subsequently, for applying the SAGE calibration, partitioning over the unknowns by manually checking the characteristics of all the sources will not be efficient. Therefore, the first challenge in utilizing the SAGE calibration to real data would be automating the partitioning over the unknowns for any given problem. Furthermore, we saw in the paper that all the mentioned calibration algorithms involve essentially  the solution of a non-linear optimization problem. Applying suitable regularization techniques using proper smoothing functions to improve the accuracy of the solutions is an issue that must be investigated further in the near future. Add to that,  that all the calibration schemes have the possibility of converging to a local optimum. Utilizing probabilistic techniques such as Simulated Annealing (SA) \citep{S.5131983} to assure that we are converging to a global optimum has the problem of decreasing the speed of convergence. Providing extra constraints for the mentioned calibration schemes which can guarantee the convergence to the real solutions could be one of the challenging areas of research in the future. Moreover, we have shown that the solver noise criterion could be used for revealing the level of accuracy in the calibrated solutions. Investigating possible systematic effects on the solver noise as well as the level of their influence are amongst the main issues for the future work.
\section*{acknowledgment}
The first author would like to \san{thank} Ernst Wit for valuable comments. The authors would like to thank P. Lampropoulos for the useful discussions along the course of this work. The first author also would like to gratefully acknowledge NWO grant 436040 and S. Z. thanks the Lady Davis foundation for support.
\bibliographystyle{mn2e}
\bibliography{references}

\begin{thebibliography}{21}
\expandafter\ifx\csname natexlab\endcsname\relax\def\natexlab#1{#1}\fi

\bibitem[{Akaike(1973)}]{H.1}
Akaike H., 1973, in Second {I}nternational {S}ymposium on {I}nformation
  {T}heory ({T}sahkadsor, 1971), Akad\'emiai Kiad\'o, Budapest, pp. 267--281

\bibitem[{Bates \& Watts(2007)}]{D.G.}
Bates D.~M., Watts D.~G., 2007, Nonlinear Regression Analysis and Its
  Applications. {Wiley-Interscience}

\bibitem[{Bilmes(1998)}]{citeulike:3334839}
Bilmes J., 1998, U.C. Berkely, TR-97-02

\bibitem[{{Boonstra} \& {van der Veen}(2003)}]{2003ITSP...51...25B}
{Boonstra} A., {van der Veen} A., 2003, IEEE Transactions on Signal Processing,
  51, 25

\bibitem[{{Born} \& {Wolf}(1999)}]{bornwolf}
{Born} M., {Wolf} E., 1999, {Principles of Optics}, {Born, M.~\& Wolf, E.}, ed.

\bibitem[{{Condon}(1974)}]{condon}
{Condon} J.~J., 1974, \apj, 188, 279

\bibitem[{Feder \& Weinstein(1988)}]{M.2}
Feder M., Weinstein E., 1988, Acoustics, Speech and Signal Processing, IEEE
  Transactions on, 36, 477

\bibitem[{{Fessler} \& {Hero}(1994)}]{J.A.1}
{Fessler} J.~A., {Hero} A.~O., 1994, IEEE Transactions on Signal Processing,
  42, 2664

\bibitem[{Gallant(1975)}]{A.R.}
Gallant A.~R., 1975, Amer. Statist., 29, 73

\bibitem[{Graves(1978)}]{L.R.T}
Graves S., 1978, British J. Philos. Sci., 29, 1

\bibitem[{{Hamaker} {et~al.}(1996){Hamaker}, {Bregman}, \& {Sault}}]{J.P.1}
{Hamaker} J.~P., {Bregman} J.~D., {Sault} R.~J., 1996, \aaps, 117, 137

\bibitem[{Kirkpatrick {et~al.}(1983)Kirkpatrick, Gelatt, \& Vecchi}]{S.5131983}
Kirkpatrick S., Gelatt C.~D. J., Vecchi M.~P., 1983, Science, 220, 671

\bibitem[{Kullback(1997)}]{K.L.D}
Kullback S., 1997, Information Theory and Statistics. Dover Publications

\bibitem[{Leshem \& van~der Veen(2000)}]{A.J.1}
Leshem A., van~der Veen A.-J., 2000, Information Theory, IEEE Transactions on,
  46, 1730

\bibitem[{Levenberg(1944)}]{K.L1}
Levenberg K., 1944, Quart. Appl. Math., 2, 164

\bibitem[{Marquardt(1963)}]{A.L.1}
Marquardt D.~W., 1963, J. Soc. Indust. Appl. Math., 11, 431

\bibitem[{{Pearson} \& {Readhead}(1984)}]{selfcal}
{Pearson} T.~J., {Readhead} A.~C.~S., 1984, \araa, 22, 97

\bibitem[{Thompson {et~al.}(2001)Thompson, Moran, \& Swenson}]{A.R.1}
Thompson A.~R., Moran J.~M., Swenson G.~W., 2001, Interferometry and Synthesis
  in Radio Astronomy, 2nd edn. {Wiley-VCH}

\bibitem[{van~der Tol {et~al.}(2007)van~der Tol, Jeffs, \& van~der Veen}]{S.1}
van~der Tol S., Jeffs B., van~der Veen A.-J., 2007, Signal Processing, IEEE
  Transactions on, 55, 4497

\bibitem[{{Wijnholds}(2010)}]{stef}
{Wijnholds} S., 2010, PhD thesis, TU Delft

\bibitem[{Yatawatta {et~al.}(2009)Yatawatta, Zaroubi, de~Bruyn, Koopmans, \&
  Noordam}]{S.2}
Yatawatta S., Zaroubi S., de~Bruyn G., Koopmans L., Noordam J., 2009, Digital
  Signal Processing Workshop and 5th IEEE Signal Processing Education Workshop,
  2009. DSP/SPE 2009. IEEE 13th, 150

\end{thebibliography}

\appendix

\section{The EM and the SAGE algorithms}\label{appen}
\subsection{EM algorithm}\label{appen1}
Considering the complete data ${\bf x}=[{\bf x}_1^T\ {\bf x}_2^T\ldots{\bf x}_K^T]^T$, where ${\bf x}_i$s are defined as (\ref{s7}), its PDF will be equal to
\begin{equation}
f_X({\bf x};{\pmb{\theta}})=\frac1{\pi ^{(KM)}|{\pmb{\Sigma}}|}\mbox{exp}\{-({\bf x}-{\bf s}({\pmb{\theta}}) )^H{\pmb{\Sigma}}^{-1}({\bf x}-{\bf s}({\pmb{\theta}}))\}.\label{a0}
\end{equation}
In (\ref{a0}) we have
\begin{equation}
{\bf s}({\pmb{\theta}})=[{\bf s}_1({\pmb{\theta}}_1)^T\ {\bf s}_2({\pmb{\theta}}_2)^T\ldots{\bf s}_K({\pmb{\theta}}_K)^T]^T,\label{e1}
\end{equation}
and
\begin{equation}
{\pmb{\Sigma}}=\begin{bmatrix}
\beta_1{\pmb{\Pi}} &      O            &\ldots &   O              \\
 O                 & \beta_2{\pmb{\Pi}}&\ldots &   O              \\
 \vdots            &       \vdots      &\ddots &\vdots            \\
 O                 &       O           &\ldots &\beta_K{\pmb{\Pi}}\\
\end{bmatrix}.\label{e2}
\end{equation}
Therefore, the log-likelihood of the complete data  ${\bf x}$ is derived from
\begin{equation}
\mbox{log}f_X({\bf x};{\pmb{\theta}})=c-\{ ( {\bf x}-{\bf s}({\pmb{\theta}}) )^H{\pmb{\Sigma}}^{-1}({\bf x}-{\bf s}({\pmb{\theta}}))\},\label{a1}
\end{equation}
where
\begin{equation*}
c=-\mbox{log}\{\pi^{(KM)}|{\pmb{\Sigma}}|\}.
\end{equation*}
Substituting (\ref{e1}) and (\ref{e2}) in (\ref{a1}), we can rewrite (\ref{a1}) as
\begin{equation}
\begin{array}{l}
\mbox{log}f_X({\bf x};{\pmb{\theta}})=c\\[5mm]
-\sum_{i=1}^K\{({\bf x}_i-{\bf s}_i({\pmb{\theta}}_i))^H(\beta_i{\pmb{\Pi}})^{-1}({\bf x_i}-{\bf s_i}({\pmb{\theta}}_i))\}.\label{a3}
\end{array}
\end{equation}
Moreover, the complete data ${\bf x}$ and the observed data ${\bf y}$ are related by the below linear transformation 
\begin{equation}
{\bf y}=[{\bf I}\ {\bf I}\ldots{\bf I}]{\bf x}={\bf{Gx}},
\end{equation}
where ${\bf G}$ is a block matrix consisting of the identity matrix ${\bf I}$ for $K$ times. Thus, they are jointly Gaussian and for a parameter value ${\pmb{\theta}}'$ we have
\begin{equation}
\widehat{{\bf x}}=\operatorname{E}\{{\bf x}|{\bf y};{\pmb{\theta}}'\}={\bf s}({\pmb{\theta}}')+\Sigma{\bf G}^H[{\bf G}\Sigma{\bf G}^H]^{-1}[{\bf y}-{\bf G}{\bf s}({\pmb{\theta}}')].\label{a4}
\end{equation}
(\ref{a4}) gives us 
\begin{equation}
\widehat{{\bf x}}_i={\bf s}_i({\pmb{\theta}}'_i)+\beta_i[{\bf y}-\sum_{l=1}^K{\bf s}_l({\pmb{\theta}}'_l)].\label{a5}
\end{equation}
as the $i$-th element of the vector $\widehat{{\bf x}}$.

In applying the EM algorithm, at $k+1$-th iteration we would like to find the parameter vector ${\pmb{\theta}}^{k+1}$ such that maximizes $\operatorname{E}\{\mbox{log}f_X({\bf x};{\pmb{\theta}})|Y={\bf y};{\pmb{\theta}}^k\}$, where ${\pmb{\theta}}^k$ is the estimation of ${\pmb{\theta}}$ obtained at $k$-th iteration. According to (\ref{a3}) it is exactly equivalent to find ${\pmb{\theta}}_i^{k+1}$ for each $i\in\{1,2,\ldots,K\}$ such that minimizes 
$\operatorname{E}\{({\bf x}_i-{\bf s}_i({\pmb{\theta}}_i))^H(\beta_i{\pmb{\Pi}})^{-1}({\bf x_i}-{\bf s_i}({\pmb{\theta}}_i))|Y={\bf y};{\pmb{\theta}}^k\}$. Therefore, at the $k+1$-th iteration of the algorithm we have\\

{\em E Step}: Calculate 
\begin{equation}
\widehat{{\bf x}}_i^k={\bf s}_i({\pmb{\theta}}_i^k)+\beta_i[{\bf y}-\sum_{l=1}^K {\bf s}_l({\pmb{\theta}}_l^k)].\label{a6}
\end{equation}  

{\em M Step}: Compute
\begin{equation}
{\pmb{\theta}}_i^{k+1}\ =\ \mbox{arg\ min}\ ||[\widehat{{\bf x}}_i^k-{\bf s}_i({\pmb{\theta}}_i)](\beta_i {\pmb{\Pi}})^{-\frac12}||^2,
\end{equation}\\[-11mm]
\begin{equation*}
 {\pmb{\theta}}_i\ \ \ \ \ \ \ \ \ \ \ \ \ \ \ \ \  \ \ \ \ \ \ \ \ \ \ \ \ \ \ \ \   \ \ \ \ \ \ 
\end{equation*}
for $i\in\{1,2,\ldots,K\}$.
\subsection{SAGE algorithm}\label{appen2}
Since in the SAGE algorithm the complete data ${\bf x}_{W_i}$ is defined as (\ref{s12}), we have 
\begin{equation}
\begin{array}{l}
\mbox{log}f_{X_{W_i}}({\bf x}_{W_i};{\pmb{\theta}}_{W_i})=-\mbox{log}\{\pi^{M}|{\pmb{\Pi}}|\}\\[5mm]
-\{ ( {\bf x}_{W_i}-\sum_{l\in W_i}{\bf s}_l({\pmb{\theta}}_{W_i}) )^H{\pmb{\Pi}}^{-1}({\bf x}_{W_i}-\sum_{l\in W_i}{\bf s}_l({\pmb{\theta}}_{W_i}))\}.\label{a7}
\end{array}
\end{equation}

At $k+1$-th iteration of the algorithm we should calculate the parameter vector ${\pmb{\theta}}_{W_i}^{k+1}$ which is maximizing $\operatorname{E}\{\mbox{log}f_{X_{W_i}}({\bf x}_{W_i};{\pmb{\theta}}_{W_i})|Y={\bf y};{\pmb{\theta}}^k\}$. Having (\ref{a7}), ${\pmb{\theta}}_{W_i}^{k+1}$ can be derived by minimizing $\operatorname{E}\{ ( {\bf x}_{W_i}-\sum_{l\in W_i}{\bf s}_l({\pmb{\theta}}_{W_i}) )^H{\pmb{\Pi}}^{-1}({\bf x}_{W_i}-\sum_{l\in W_i}{\bf s}_l({\pmb{\theta}}_{W_i}))|Y={\bf y};{\pmb{\theta}}^k\}$ with respect to ${\pmb{\theta}}_{W_i}$. Therefore, the SAGE algorithm's steps at the $k+1$-th iteration would be written as \\

{\em {SAGE E Step}}: Calculate 
\begin{equation}
\widehat{{\bf x}}_{W_i}^k=\operatorname{E}\{{\bf x}_{W_i}|{\bf y},{\pmb{\theta}}^k\}={\bf y}-\sum_{\substack{j=1\\j\neq i}}^{m}\sum_{l\in {W_j}}{\bf s}_l({\pmb{\theta}}_{W_j}^k),\label{a8}
\end{equation} 
which is derived from (\ref{s14}).\\\\

{\em SAGE M Step}: Compute
\begin{equation}
{\pmb{\theta}}_{W_i}^{k+1}\ =\ \mbox{arg\ min}\ ||[\widehat{{\bf x}}_{W_i}^k-\sum_{l\in {W_i}}{\bf s}_l({\pmb{\theta}}_{W_i})]({\pmb{\Pi}})^{-\frac12}||^2,
\end{equation}\\[-12mm]
\begin{equation*}
 {\pmb{\theta}}_{W_i}\ \ \ \ \ \ \ \ \ \ \ \ \ \ \ \ \  \ \ \ \ \ \ \ \ \ \ \ \ \ \ \ \   \ \ \ \ \ \ \ \ \ \ \ \ \ \ \ 
\end{equation*}
for $i\in\{1,2,\ldots,K\}$.

The EM and in particular the SAGE algorithms have the well-known advantage that they ensure that the likelihood gets increased at each iteration step.  However, they suffer from two negative features. Their rate of convergence is exponential, and they cannot guarantee the convergence to a global optimum.  Nevertheless, they are significantly benefited from choosing a suitable starting point which can be the topic of the future work. 

\bsp
\label{lastpage}
 \end{document}